\documentclass[preprint2]{aastex61}
\usepackage{amsmath,amstext}
\usepackage[T1]{fontenc}
\usepackage{graphics,graphicx}

\def\etal{{et\,al.}}
\def\msun{M$_{\odot}$}
\def\fd{\hbox{$.\!\!^{\rm d}$}}            
\def\farcs{\hbox{$.\!\!^{\prime\prime}$}}  
\def\degs{\ifmmode ^{\circ}\else$^{\circ}$\fi}
\def\amin{\ifmmode ^{\prime}\else$^{\prime}$\fi}
\def\asec{\ifmmode ^{\prime\prime}\else$^{\prime\prime}$\fi}
\newbox\grsign \setbox\grsign=\hbox{$>$}
\newdimen\grdimen \grdimen=\ht\grsign
\newbox\laxbox \newbox\gaxbox
\setbox\gaxbox=\hbox{\raise.5ex\hbox{$>$}\llap
     {\lower.5ex\hbox{$\sim$}}}\ht1=\grdimen\dp1=0pt
\setbox\laxbox=\hbox{\raise.5ex\hbox{$<$}\llap
     {\lower.5ex\hbox{$\sim$}}}\ht2=\grdimen\dp2=0pt
\def\gax{$\mathrel{\copy\gaxbox}$}

\newcommand{\up}{\mbox{$u^\prime$}}
\newcommand{\gp}{\mbox{$g^\prime$}}
\newcommand{\rp}{\mbox{$r^\prime$}}
\newcommand{\ip}{\mbox{$i^\prime$}}
\newcommand{\zp}{\mbox{$z^\prime$}}

\begin{document}

\title{The benefit of simultaneous seven-filter imaging: \\
10 years of GROND observations\thanks{Based on observations made with 
ESO Telescopes at the La Silla Paranal Observatory under programme 
ID 292.D-5029(A).}}

\author{J. Greiner
}
\affil{Max-Planck-Institut f\"ur extraterrestrische Physik, 85740 Garching,
    Germany}
\email{jcg@mpe.mpg.de}

\begin{abstract}
A variety of scientific results have been achieved
over the last 10 years with the GROND simultaneous 7-channel imager at the 
2.2m telescope of the Max-Planck Society at ESO/La Silla. While designed
primarily for rapid observations of gamma-ray burst afterglows, the
combination of simultaneous imaging in the Sloan $g'r'i'z'$ and 
near-infrared $JHK_s$ bands at a medium-sized (2.2\,m) telescope and 
the very flexible scheduling possibility has resulted in an
extensive use for many other astrophysical research topics,
from exoplanets and accreting binaries to galaxies and quasars.
\end{abstract}


\section{Introduction}

An increasing number of scientific questions require the
measurement of spatially and spectrally resolved intensities
of radiation from astrophysical objects. Over the last decade,
transient and time-variable sources are increasingly moving
in the focus of present-day research (with its separate naming of
``time-domain astronomy''), recently boosted spectacularly by the
follow-up of gravitational wave sources.
If the spatial scale of such a study is small (few arcmin), integral field
spectrographs such as PMAS (3.6m Calar Alto)
or MUSE (VLT) or ESI/OSIRIS (Keck) are the 
instruments of choice.
If crowding is not an issue, (objective) prism spectroscopy
is a valuable option \citep{Teplitz+2000}.
For large scales, simultaneous multi-channel imaging is applied.
The physical measurement goals often request a compromise 
between spatial, temporal or spectral resolution, which adds to the
challenges of the measurement principle.

Simultaneous imaging in different filter-bands 
(whether Johnson $UBVRIJHK$ or Sloan \up\gp\rp\ip\zp\ or anything else)
is of interest
in a variety of astrophysical themes. The primary aim is to
measure the spectral energy distribution (SED) or its evolution in variable 
astrophysical objects, in order to uncover the underlying emission mechanism.
Examples are, among others,
(1) monitoring of all kinds of variable stars (flare stars, cataclysmic
  variables, X-ray binaries) to determine the outburst mechanisms and
  differentiate between  physical state changes and changes induced
  by geometrical variations, like eclipses;
(2) follow-up of gamma-ray burst (GRB) afterglows for e.g. rapid
  redshift estimates, mapping the SED evolution 
  to measure circumburst parameters, or the search for dust destruction;
(3) monitoring of AGN to understand the physical origin of the
  observed variability;
(4) determining the inclination of X-ray heated binaries \citep{orosz}; 
(5) mapping of galaxies to study their stellar population;
(6) multi-color light curves of supernovae to, e.g., recognize
dust formation \citep{tpm06}; 
(7) differentiating achromatic microlensing events \citep{pac86}
   from other variables with similar light curves;
(8) identifying objects with peculiar SEDs, e.g. 
  photometric redshift surveys for high-$z$ active galactic nuclei, or
  identifying brown dwarfs;
(9) observations of transiting extrasolar planets to infer orbital periods,
  multiplicity of planets, or characteristics
  of their atmospheres \citep{jha00}; or
(10) mapping of reflectance of solar system bodies as a function of 
their rotation to map their surface chemical composition \citep{jew02}.

Instruments with simultaneous imaging capability in different filter bands
prior to the GROND development include 
ANDICAM \citep{dep98},
BUSCA \citep{rei99},
HIPO \citep{dun04},
MITSuME \citep{kky07},
TRISPEC \citep{wny05},
SQIID \citep{edf92},
and ULTRACAM \citep{dhi07}.
GROND-inspired instruments include the 6-channel RATIR \citep{Butler2012} 
and the 4-channel ROS2 \citep{Spano2010} instruments.
Further projects for simultaneous multi-band instruments are the 8-channel
OCTOCAM \citep{Gorosabel2010}, selected as part of the Gemini instrumentation 
program in 2017 \citep{Roming2018}, the 4-channel 
SPARC4 \citep{Rodrigues+2012} planned for installation at the 1.6 m telescope 
of the Pico dos Dias Observatory (Brazil) \citep{Bernardes2018}, 
an unnamed 8-channel imager for the IRTF \citep{Connelley2013},
and the SIOUX project \citep{Christille+2016}.
In comparison, the GROND instrument \citep{Greiner+2008}
at the 2.2\,m telescope of the Max-Planck Society (MPG) in La Silla 
(ESO/Chile) with its 7 simultaneous channels so far still delivers
the largest degree of multiplexing at such a telescope size.

\begin{figure}[th]
\vspace{-3.5cm}
\includegraphics[width=1.0\columnwidth]{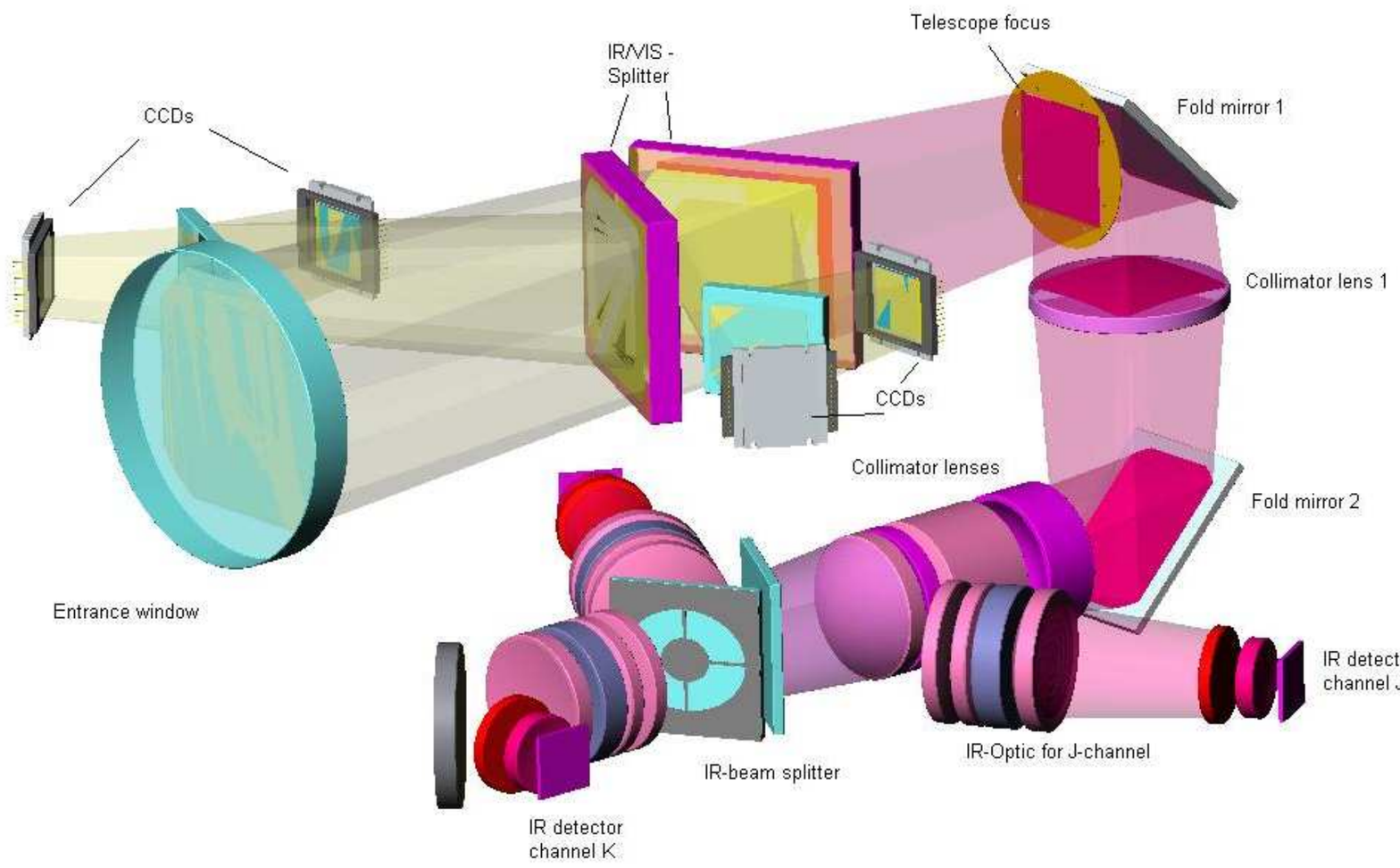}
\vspace{-3.5cm}
\caption{Scheme of the optical beam path of GROND with the optical components 
and the detectors labeled. [From \cite{Greiner+2008}]
\textcopyright AAS. Reproduced with permission.
\label{3Dopt}}
\end{figure}

After a short description of the main features of the instrument
and operational aspects ($\S 2$), I describe some of our prime 
scientific results obtained via GROND observations, foremost
for GRBs ($\S 3$) and transients ($\S 4$), but also
other science topics where color information on short timescales
is important ($\S 5-9$). While this is predominantly a review, 
it contains hitherto unpublished results, e.g. on the discovery of a
hitherto unknown T5 brown dwarf.

\begin{figure*}[t]
\includegraphics[width=13.0cm]{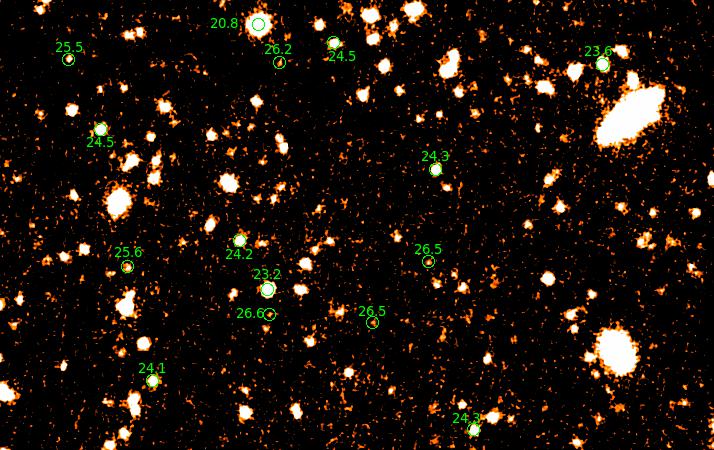}\hfill\parbox[t]{4.2cm}{\vspace*{-7.0cm}\hspace*{7.cm}\caption{An example for the sensitivity of GROND@2.2m,
reaching \gp = 26.5 mag in 3.5 hrs exposure time \citep{Yates2015},
likely one of the deepest images from a ground-based 2\,m class telescope.
The green numbers are SDSS-calibrated \gp-magnitudes.
This \gp-band image is 1\farcm1 $\times$ 1\farcm8; North is up, and East to the left.
\vspace{0.6cm}
}}
\label{deepfield}
\end{figure*}

\section{The GROND instrument and its operation}

The primary goal was to rapidly identify GRB afterglows and measuring their 
redshift. This led to the concept of a camera which allows simultaneous
observations in multiple
filters throughout the optical and near-infrared region.
The simultaneity is dictated by the fact that a typical GRB afterglow initially
fades by about 2--3 mag within 5--10 min after the GRB, and by another 3 mag
in the following 50 min, thus rendering cycling through different filters
useless.
Furthermore, with the advent of {\it Swift}'s detection of $\sim$100 GRBs/yr, 
follow-up of each GRB with an 8\,m telescope became impractical, and 
some knowledge-based pre-selection was needed.
Four bands were implemented in the visual, 
plus three (standard $JHK_s$) bands in the near-infrared (NIR).
The separation of the  different photometric bands was
achieved using dichroic beamsplitters (in the converging beam), 
whereby the short wavelength part of the light is
always reflected off the dichroic, while the long-wavelength part
passes through (Fig. \ref{3Dopt}).
The use of dichroics implies that adjacent bands do have identical 50\%
transmission wavelengths, making the Sloan filter 
system \citep{fig96} the obvious choice for the visual bands.

The field-of-view (FOV) of the camera was designed, on one hand, to cover
the typical few arcmin extent of GRB error boxes, 
and on the other hand have a pixel scale less than the mean seeing  
to allow for accurate photometry. 
Mounted at the MPG-owned 2.2\,m diameter f/8 telescope on La Silla (ESO/Chile) 
with an intrinsic image quality of 0\farcs4,
the FOV of each visual band is $5.4\,\times\,5.4$ arcmin$^2$,
(2048x2048 CCD with plate scale 0\farcs158/pixel),
and $10\,\times\,10$ arcmin$^2$ in the NIR using a focal reducer 
($1024\,\times\,1024$ Rockwell HAWAII-1 array with a plate scale of 
0\farcs60/pixel).
A Sumitomo closed-cycle cooler provides a temperature of 65\,K for the
NIR detectors and 80\,K for the focal reducer optics, with simple 
damping preventing any telescope/instrument vibrations which could 
degrade the image quality. The best GROND images have
a full-width-half-maximum of 0\farcs6, dominated by the dome seeing.
This allows us to linearly increase sensitivity by adding more exposure
(stacking) up to 3--4 hrs, before becoming background-dominated 
(see, e.g. Fig. \ref{deepfield}).

The standard detector readout systems which were used at ESO at the time 
were implemented, i.e. FIERA \citep{Beletic1998} for the visual channels, 
and IRACE \citep{Meyer1998} for the NIR channels.
This makes for a very flexible readout scheme, where e.g. NIR exposures
continue during the CCD-readout. Since the $JHK_s$ channels operate 
fully synchronously, a 10 s exposure was adopted as a compromise
between not saturating the $K_s$-band while maximizing $J$-band exposure
per telescope dither position. In addition,
a separate internal dither mechanism was implemented in the $K_s$-band;
full details can be found in \cite{Greiner+2008}.

While originally foreseen to only operate in robotic target-of-opportunity 
mode for chasing GRB afterglows, the
GROND operation scheme was designed flexibly enough to allow also
visitor-mode style ``manual'' observations.
All parameters for GROND observations 
can be adapted through standard ESO-style observation blocks (OBs)
which are used for
all observations, whether visitor/service mode or robotic.
Normal observing program observers use the canonical p2pp tool 
(P2PP Manual 2007\footnote{see 
www.eso.org/observing/p2pp/P2PP-tool.html\#Manual}),
while in the case of GRB observations, OBs are generated in real-time 
by an automatic process. 
A special commandable mirror allows to switch between GROND and the
other two 2.2\,m instruments (WFI, FEROS) within 20 s.
At the start of each OB, the instrument is automatically focused by moving
the telescope's secondary mirror. 

Best possible instrument efficiency has been the main driver 
during the design and development of GROND. As a result,
the total efficiency in the visual bands
is about 70\% (except the $z'$ band), and is still above 50\% for the
three NIR bands \citep{Greiner+2008},
despite the eleven lenses per channel and the comparatively low quantum
efficiency of the 2001-built HAWAII detectors.
Thus, even in a single filter, GROND is the most sensitive instrument
at a 2\,m-class telescope. Due to the simultaneous imaging in
7 channels, GROND is likely the instrument with the highest 
photon-detection efficiency in the 0.4--2.5 $\mu$m band.

GROND was commissioned at the MPG 2.2m telescope at La Silla (ESO, Chile)
in April/May 2007,
and the first gamma-ray burst followed up was GRB 070521 \citep{gck07a}.
For the first few months (until end of September 2007), follow-up observations
depended on the willingness of the scheduled observers to share observing time.
Thereafter, weather permitting, a general override permission and a 15\% share
of total telescope time allowed us to follow every well-localized
GRB which was visible from La Silla, with only few exceptions.
A MPE directorial decision terminated this systematic GRB follow-up program 
with GROND at the beginning of October 2016.

\begin{figure}[t]
\vspace{-0.2cm}
\hspace{-1.3cm}\includegraphics[angle=270,width=10.7cm]{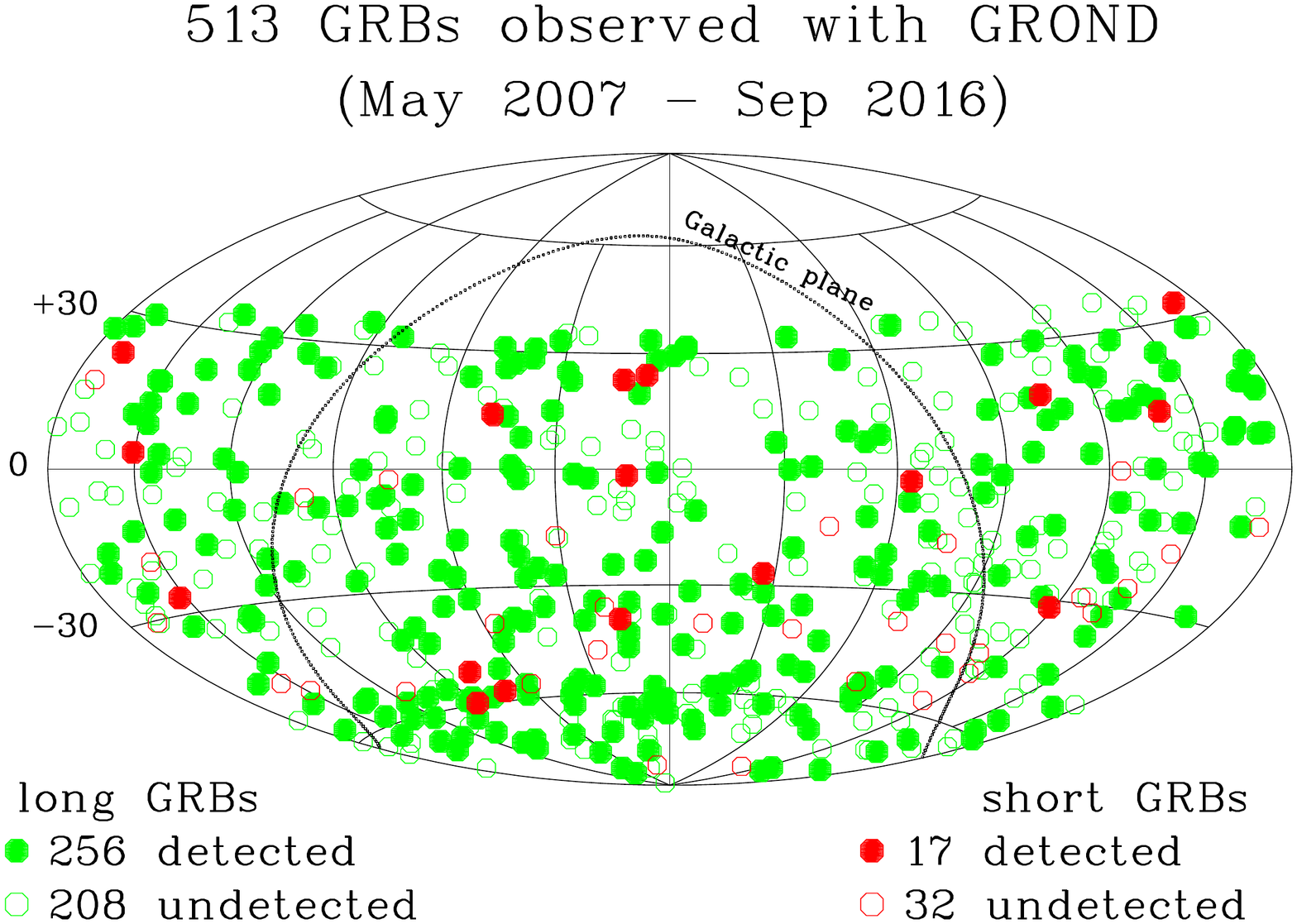}
\vspace{-1.0cm}
\caption{Sky distribution in equatorial coordinates of the GROND GRB
sample.}
\label{grondGRBs}
\end{figure}

\section{Gamma-ray bursts}

\subsection{Long- and short-duration GRBs}

GRBs are the most luminous electromagnetic sources on the sky, 
releasing in less than a minute the energy output of the Sun over 
its entire life.  
GRBs form two sub-groups according to their duration:
(i) Long-duration GRBs ($>$2 s) are firmly linked to the collapse 
of massive stars \citep{Hjorth03, Stanek03},
thus probing sites of star formation with little delay, as the star's 
lifetimes are measured in megayears. GRBs have been seen
up to the highest measured redshifts.
(ii) Short-duration GRBs are commonly believed to originate
from the merging of compact stars, as verified by the recent detection of
gravitational waves from GRB 170817A \citep{LIGO_APJL_MM}.

Present $\gamma$-ray instrumentation provides a
detection rate of about one GRB per day, and thus GRBs act as 
frequently available signposts throughout the Universe.
Over  the last 2 decades,
these ultra-luminous cosmological  explosions have been transformed from a
mere curiosity  to essential tools for the study of high-redshift
stars and galaxies, early structure formation and the evolution of
chemical elements.

\subsection{GROND Observing statistics}

A total of 842 GRBs were promptly localized by the Neil Gehrels {\it Swift} 
Observatory  and 301 by other missions\footnote{see 
http://mpe.mpg.de/$^\sim$jcg/grbgen.html for a complete list} 
(the majority with
error boxes much larger than the 10\amin\ of GROND) between May 2007 and
September 2016.  879 of these happened at declination smaller than +36\fd5 
(which is about the northern-most declination reachable with GROND due to
a minimum 20\degs\ horizon distance requirement of the 2.2m telescope),
out of which 513 were followed-up with GROND. 
256 of the 464 long-duration GRBs were detected, and 17 of the 49 
short-duration GRBs (Fig. \ref{grondGRBs}).
For the subset of 709 Swift-detected GRBs with immediately (up to few hours) 
well-localized {\it Swift}/XRT afterglow positions, 532 were observable
for GROND, and 439 were actually observed. This implies a follow-up
efficiency of these well-localized {\it Swift}-GRBs of 82\%, with bad weather
periods and main-mirror coating events being the largest impact factors 
among the not-observed sources.

\subsection{GRBs as high-redshift probes}

\begin{figure}[t]
\hspace{-0.1cm}\includegraphics[width=8.6cm]{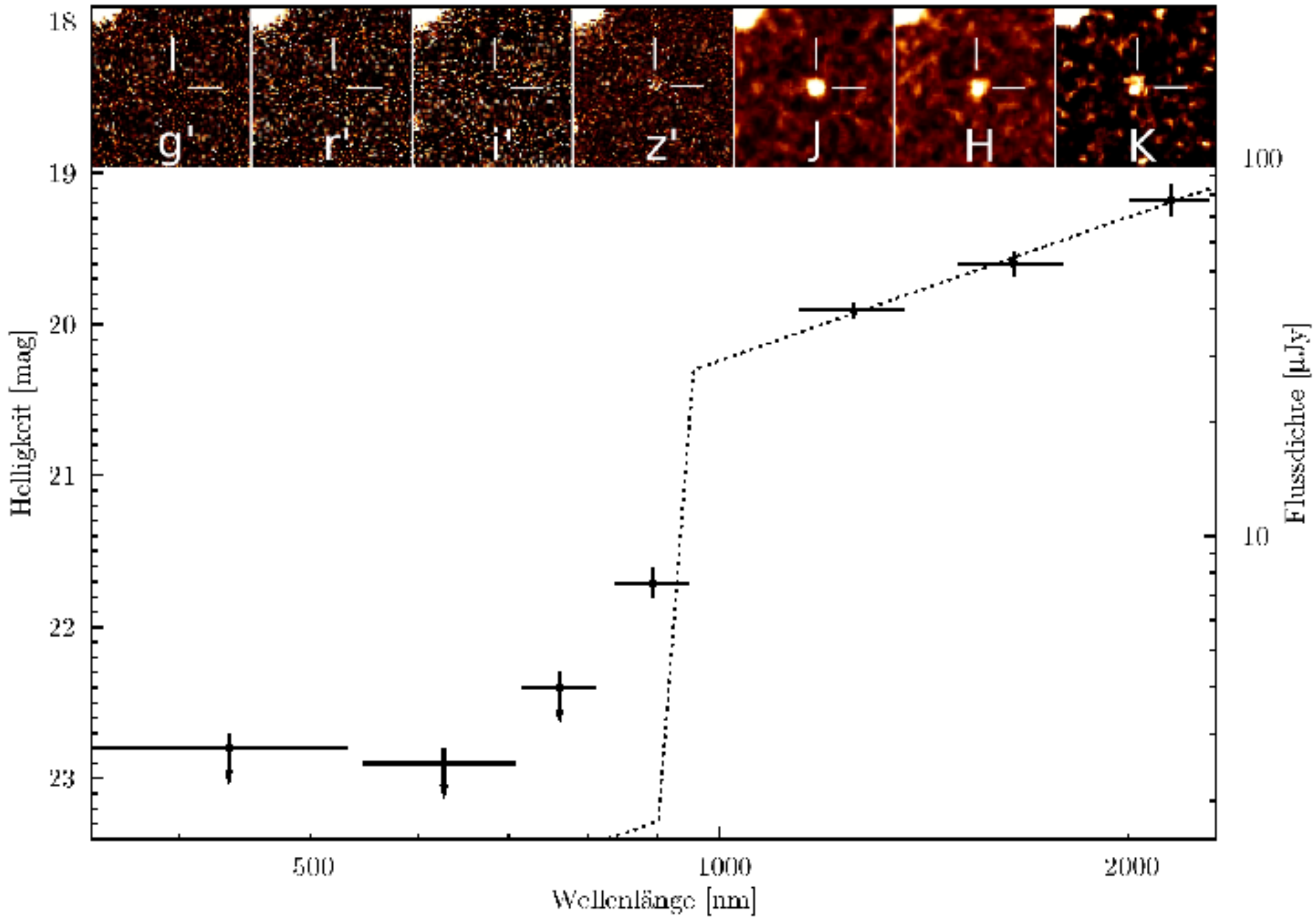}
\vspace{-0.55cm}
\caption{GROND spectral energy distribution and the corresponding
image cut-outs (top row) for GRB 080913 at z=6.7 \citep{gkf09}.
The GROND observation started about 6 min after the Swift/BAT trigger, and
the photometric redshift was available 35 min after the trigger,
formed from the stack of the first 3 OBs. This information 
was used to trigger FORS spectroscopy at ESO/VLT which 
confirmed the GROND photo-$z$ to be accurate to within 5\% [From \cite{gkf09}].
\textcopyright AAS. Reproduced with permission.
}
\label{grb080913}
\end{figure}

GROND operations started very promising: about 1 year after
commissioning, the afterglow of GRB 080913 at z=6.7 (see Fig. \ref{grb080913})
was discovered with GROND \citep{gkf09},
and spectroscopically confirmed with ESO/VLT spectroscopy. This 
served as the ``proof-of-concept'' for using simultaneous 
multi-band photometry of GRB afterglows to accurately measure
photometric redshifts. Later on, GROND significantly contributed to the 
record-braking GRBs 090423 \citep{Tanvir2009} and 090429B 
\citep{Cucchiara+2011} by providing data for additional filters
or allowing to calibrate the typically small field-of-view
NIR instruments on the 8-10m telescopes.

Among 273  GRB afterglows detected with GROND, we have not found a single
GRB afterglow with only a $HK_s$ detection, i.e. a $J$-band drop-out.
Unless $z>10$ GRBs are systematically underluminous, and thus below the
GROND threshold ($H(AB)>21$ mag, $K_S(AB)>20$ mag in one hour exposure),
the relative frequency of such $z>10$ GRBs is below 0.4\% (1$\sigma$).
This is consistent with recent predictions of the redshift distribution 
of GRBs, similar to our earlier estimate \citep{gkk11}
of $\sim$5\% of GRBs at $z>5$ \citep[e.g.][]{Elliott+2012, Le+Mehta17}.

\subsection{Dust and Dark GRBs}

Soon after the discovery of GRB afterglows it became clear that
the detection rate in the optical wavelength range was substantially
lower than that in X-rays \citep[e.g.][]{ggp98, ksm00, dfk01}.
The reasons for the occurrence of such ``dark'' bursts were  
first discussed systematically in \cite{fyn01} and \cite{lcg02}.
These involve
(i) either an intrinsically low luminosity, e.g. 
  an optically bright vs. optically dark dichotomy, 
or (ii) a large extinction by intervening material, either very
  locally around the GRB, or along the line-of-sight through the host galaxy,
or (iii) high redshift ($z>5-6$), so that Ly$\alpha$
  blanketing and absorption by the intergalactic medium
  would prohibit detection in the frequently used $R$ band \citep{lar00}.
A sample of 39 long-duration GRB afterglows,
complete in observational bias and redshift, and 
observed with GROND within 4 hrs, established the fraction of dark bursts 
to be 18$\pm8$\%. Among these dark bursts, the different shape of the
spectral energy distribution allows us to differentiate between
two options:
$57\pm14$\% are due to moderate dust extinction enhanced due to moderate
redshift, while
$28\pm14$\% are due to flux depression because of high redshift, $z>5$
\citep{gkk11}. Since the afterglow detection rate of this sample was very 
high (92\%; just three GRB afterglows missed) the above errors also include
potential intrinsically faint GRBs, where the maximum brightness
during the first 4 hrs after the GRB was below the sensitivity threshold
of GROND@2.2m.

\begin{figure}[t]
\includegraphics[width=8.4cm]{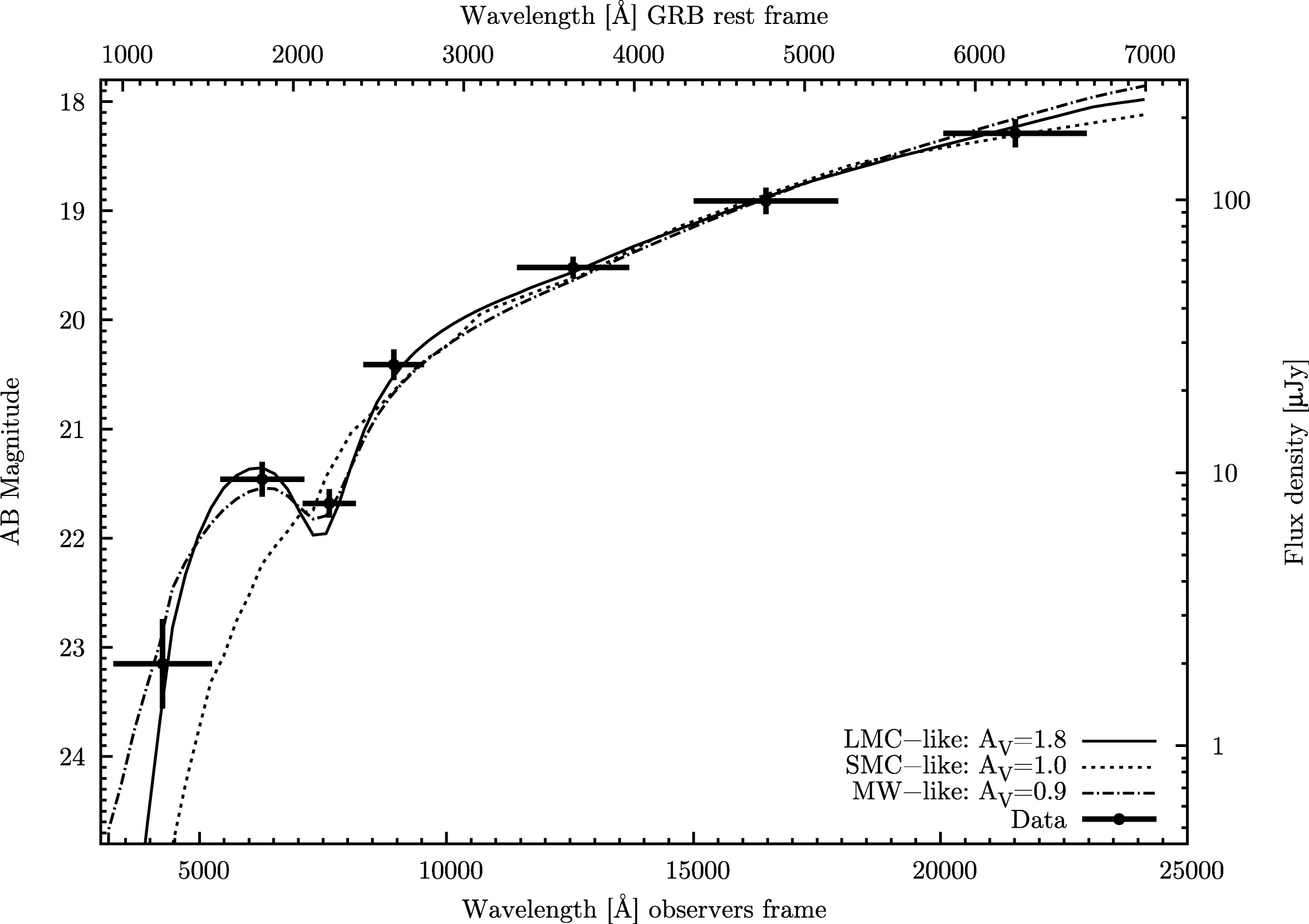}
\caption{GROND \gp\rp\ip\zp$JHK_s$ spectral energy distribution of 
the afterglow of GRB 070802, showing a clear drop of the \ip-band, interpreted
as the redshifted 2175 \AA\ bump in the GRB host galaxy [From \cite{kkg08}].
\textcopyright AAS. Reproduced with permission.
}
\label{grb070802}
\end{figure}

Another early and surprising result was the very dusty GRB 070802
at a redshift of $z=2.45$. The SED deviated clearly from 
the typical synchrotron power law shape, showing increasing curvature 
towards the bluest band and a low-flux 'outlier' in the \ip-band
(Fig. \ref{grb070802}). 
We interpreted the \ip-band drop as extinction by the 2175 \AA\ feature,
redshifted  in the GRB host galaxy \citep{kkg08}.
This was one of the first and clearest detections of the 2175 \AA\
feature at high redshift, and was later confirmed by optical
spectroscopy \citep{Eliasdottir+2009} with VLT/X-shooter.

Motivated by these examples of strong host-intrinsic extinction
and the availability of our unique GROND sample of dusty GRBs, 
a search for \citep{Rossi+2012},
or detailed analysis \citep{kgs11} of their host galaxies has been undertaken.
This revealed systematic differences in their properties relative to the
hosts of optically bright GRBs: they are systematically redder,
more luminous and more massive,
suggesting chemically evolved hosts \citep{kgs11}.
This finding established that the dust along the sight-line of GRBs
is often related to global host properties, and thus not located in 
the immediate GRB environment as expected for a massive star dying
within its star forming region. 
By now, this correlation is well accepted, and used in an inverted way
to search for infrared-bright host
galaxies of GRBs without optical afterglow, in order to obtain redshifts and 
host details 
\citep{Chrimes18}.
This would help to understand whether the aversion of long-duration GRBs with 
bright optical afterglows to massive, luminous galaxies is indeed a
generic metallicity bias \citep{Fruchter+2006, Graham+Fruchter2017}, 
or largely a selection effect.

\subsection{Fireball model tests}

Afterglow emission from GRBs was predicted 
\citep{PaczynskiRhoads1993, Katz1994, mer97, SariPiran97}
prior to its discovery with BeppoSAX \citep{Costa1997, Paradijs1997}.
This afterglow emission is commonly described with the fireball model
\citep{mer97}. When the relativistically expanding blast wave interacts 
with the circumburst medium, an external shock is formed in which  
relativistic electrons gyrating in  magnetic fields radiate synchrotron 
emission \cite{wij97, wig99}. Implicitly assuming
that the electrons are ``Fermi'' accelerated at the relativistic shocks,
and that they have a power-law
distribution with an index $p$, their dynamics can be expressed
with the following 4 parameters:
(1) the total internal energy in the shocked region released in the
   explosion,
(2) the density $n$ (and its radial profile) of the surrounding medium,
(3) the fraction of shock energy that goes into electrons, $\epsilon_e$,
(4) ratio of the magnetic field energy density to the total thermal energy,
$\epsilon_B$.
This minimal and simplest afterglow model has only five 
parameters (not counting the distance/redshift).

The evolution of the afterglow emission in frequency space and with time
depends on a number of additional boundary conditions, such as the 
properties of the burst environment
(e.g., radial gas density profile, dust),
on the progenitor (e.g., temporal energy injection profile),
and details of the shock. 
Measuring the energetics (the fraction of energy going into the electrons
$\epsilon_e$ or into the magnetic field $\epsilon_B$)
or the energy partition ($\epsilon_e$/$\epsilon_B$) 
has been challenging over the last 20 years.
One particular difficulty is to distinguish between the fast or slow cooling
stage which introduces an ambiguity in the explanation of 
the spectrum in terms of the physical model parameters.
The degeneracy between several of the above parameters makes it even more
difficult to draw astrophysical conclusions from a given data set.
Thus, many previous attempts in testing the fireball scenario
had to make compromises, i.e. make assumptions about individual parameters 
\citep[e.g.,][]{PanaiKumar02, Yost03, Chandra2008, Cenko10, 
gkn13, Laskar14, Varela+2016}.
Contradictions between results based on analyses with different assumptions
surfaced only in the rare cases where the same 
GRB afterglows were analyzed using different data sets
\citep[e.g.,][]{mkr10, Cenko11}.

\begin{figure}[t]
\vspace{-1.0cm}
\hspace{-0.9cm}\includegraphics[width=10.cm]{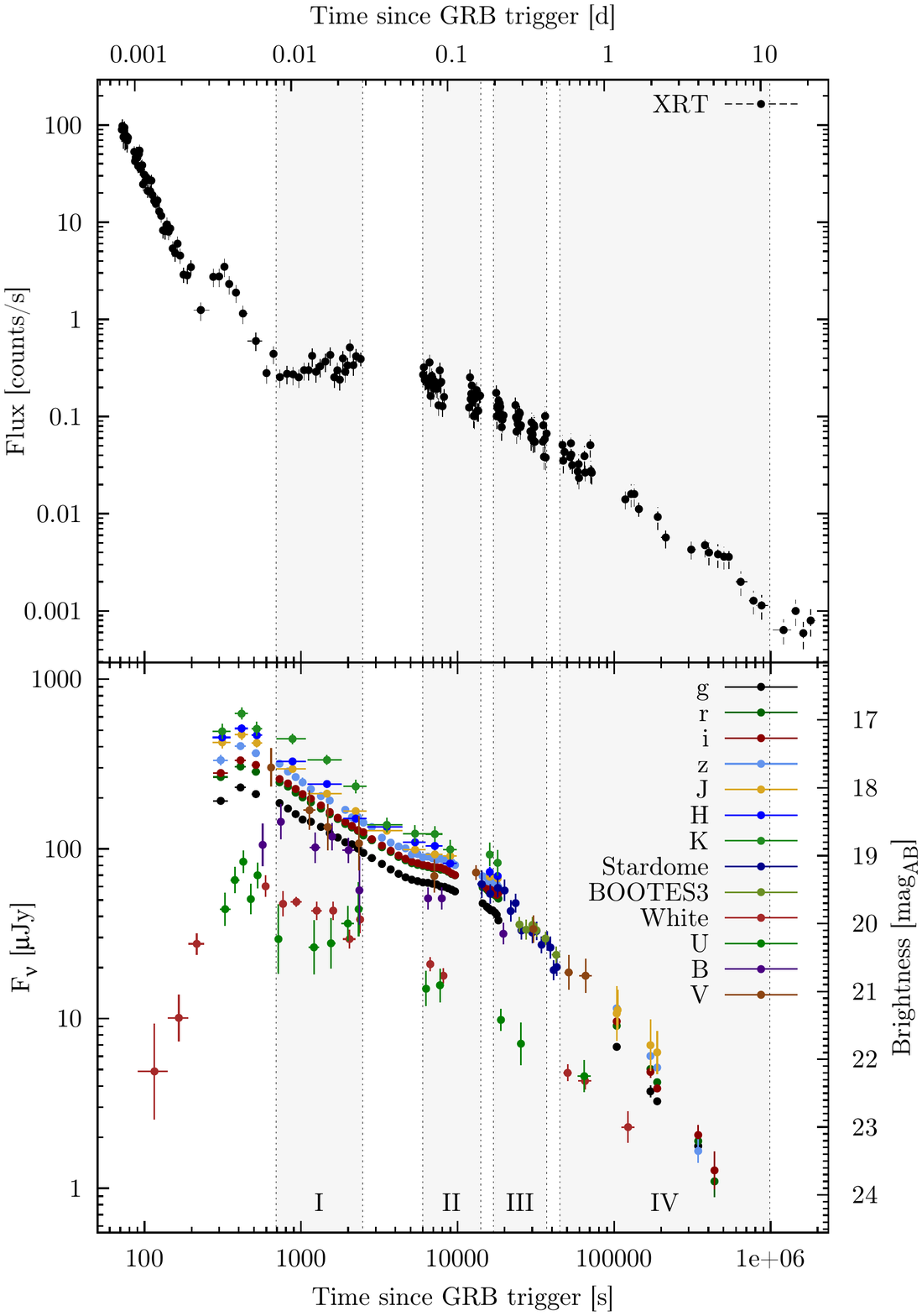}
\vspace{-1.5cm}
\caption{Light curve of the X-ray (top panel) and UV-to-NIR (bottom panel)
afterglow of GRB 091029. Grey regions show the time intervals for which
broad-band SEDs were formed. The nearly complete decoupling of the light curves
in the two panels is difficult to reconcile with the fireball scenario.
[From \cite{fgs12}].
}
\label{grb091029}
\end{figure}

With its seven simultaneous channels, GROND provides an obvious advantage 
in these studies. Consequently, a number of attempts have been made
to obtain data sets which would allow us to derive  conclusions
on the fireball parameters. The results are somewhat mixed, despite the fact
that for most of these GRBs we achieved full wavelength coverage down to
the sub-millimeter and radio bands. One of these unsatisfying examples is the
bright afterglow of GRB 100621A. Three different emission components 
were identified, each with 
different spectral slope and temporal evolution, making a solution
of even the simplest fireball scenario impossible \citep{gkn13}.

In a number of cases, the data collected with GROND (in conjunction with
{\it Swift}/XRT and the long-wavelength coverage from the sub-mm to the radio)
demonstrate convincingly that the most simplistic fireball scenario
does not describe the data well, and thus extensions are required.
GRB 091127 \citep{fgs11} was likely the first GRB afterglow
with clear evidence for a moving cooling break, as expected in the fireball 
model, and even a measurement of the sharpness of the cooling break. 
However, the temporal evolution of the cooling break was clearly inconsistent 
with the standard fireball scenario.
As one possible explanation of the data set, a temporal dependence of 
$\epsilon_B$ was proposed \citep{fgs11}, though a theoretical motivation
remains to be given.
The case of GRB 091029 may be extreme, with completely
 decoupled optical/X-ray behaviour (Fig. \ref{grb091029}). A non-standard
assumption for at least three fireball parameters was necessary,
i.e.  only a 2-component model with separate evolutionary states of each
component could potentially explain the data set \citep{fgs12}.  In several 
other GRB afterglows, the slopes of the electron distribution $p$ as derived
from the spectral slopes are
clearly $<$2, as opposed to the canonical $p = 2.2-2.3$, thus leaving
the particle acceleration mechanism and the high-energy cut-off(s) unsolved.
Obviously, such results are unsatisfactory, and fresh ideas are needed
to understand these events.

On a more positive note, GRB 121024A \citep{Varela+2016} shows a
multi-colour light curve which is similar in X-rays and the 
optical/NIR band, and has additional sub-mm and radio data. This
provided a showcase for an explanation within the basic fireball 
scenario. The 'grain of salt' was that some of the fireball parameters
had rather extreme values, outside the range normally anticipated
(though we might be mislead by our expectations).
So far the best, though still not perfect, case was GRB 151027B.
Combining the X-ray and optical/NIR measurements with radio and ALMA data
we could solve the fireball system,  except for one parameter-pair 
ambiguity.
Adopting the lowest-allowed total energy, all fireball 
parameters are well constrained, to at least  a factor of three.
The surprisingly and yet unexplained strong variability
of the radio emission meant that those data were  unusable in the fireball
analysis \citep{gbw18}. This prevented a full-fledged test of the
basic fireball scenario, including its temporal evolution. Interestingly, 
GRB 160625B also shows such strong radio variability \citep{Alexander+2017}, 
suggesting that care
must be exercised when using sparse radio data in GRB fireball modelling.

\subsection{From prompt to afterglow emission: 
flares, bumps and jumps}

One of the main motivations for building GROND was the fact that
both the observed early-time rise/decay as well as non-powerlaw ``bumps''
(e.g. in GRB 021004; \citealt{Lazzati2002}) exhibit variations faster
than the time it takes to cycle through a number of filters.
Only systematic observations in different filters as synchronous as possible
can overcome the ambiguity between effects of a changing spectrum or a 
highly variable achromatic emission.

\begin{figure}[t]
\includegraphics[width=8.7cm]{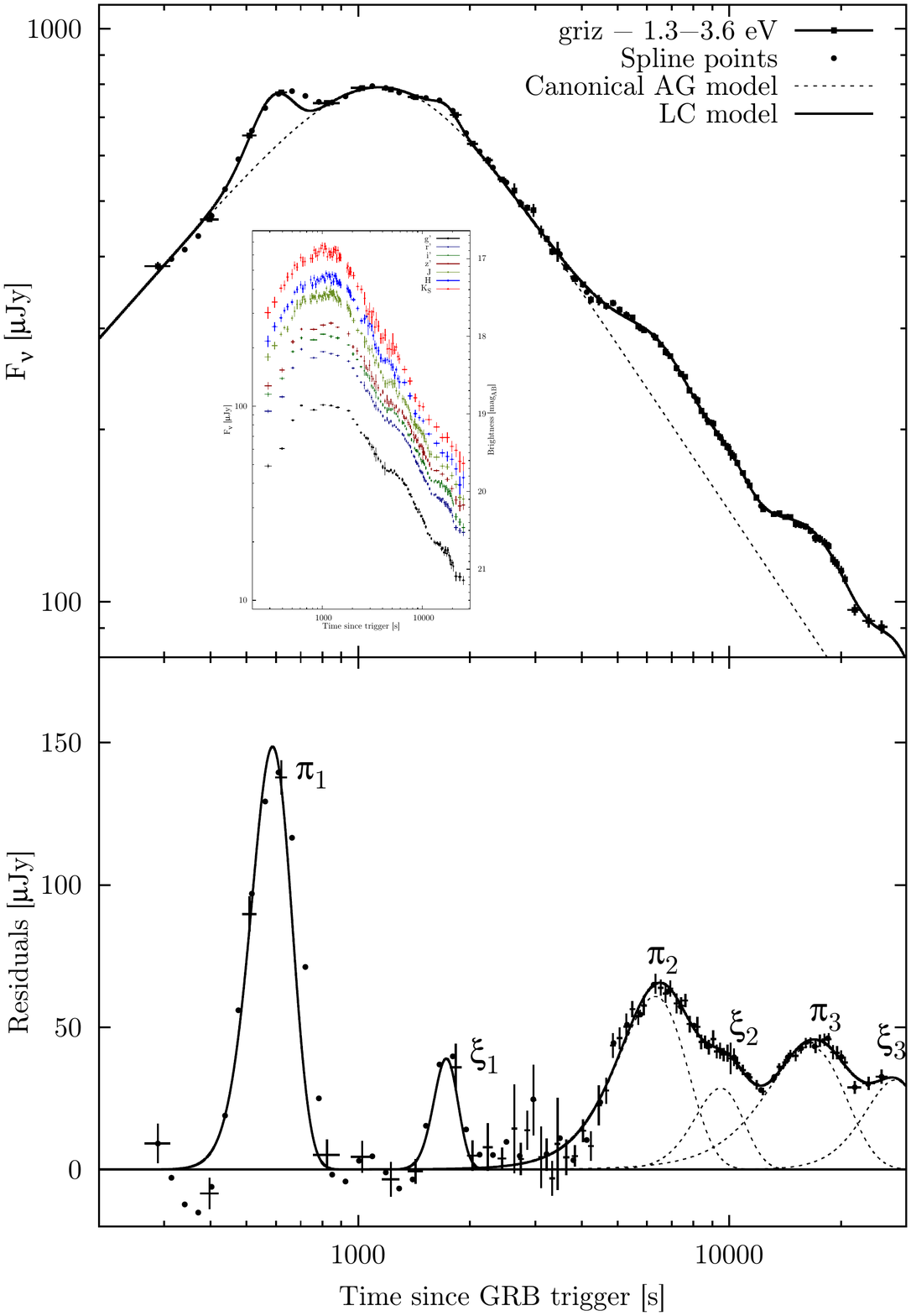}
\vspace{-1.cm}
\caption{White light curve (upper panel) of the afterglow of GRB 071031,
with the inset showing the complete 547 GROND \gp\rp\ip\zp$JHK_s$ data points.
The data were fitted using the sum 
of a smoothly connected power law for the canonical afterglow (dashed line) 
and Gaussian profiles to account for the evident flux excesses (solid line). 
The lower panel shows the residuals to the smoothly connected power law, 
and the six Gaussian models.
[From \cite{kgm09}]. \textcopyright AAS. Reproduced with permission.
}
\label{grb071031}
\end{figure}

Previously, four different mechanisms have been proposed to reproduce
bumps or flares in optical afterglow light curves:
(i) a superimposed reverse shock component for early flares, 
(ii) inhomogeneities in the circumburst medium \citep[e.g.,][]{Wang+Loeb2000},
or (iii) the angular distribution of the energy in the jet (patchy shell model; e.g.,
\citealt{Kumar+Piran2000}),  or (iv) late energy injection by refreshed
shocks \cite[e.g.,][]{Rees+Meszaros1998} for later flares. However, a clear 
discrimination in the few individual, previously studied cases was not 
possible due to the lack of 
broad-band spectral information.

A first exciting case to demonstrate the advantage of the GROND seven-band
imaging was GRB 071031: 
Superimposed onto the canonical afterglow emission, we found bumps which have a 
harder SED and appear to be similtaneous in the optical/NIR and at X-rays.
Although emission from external shocks or a combination 
of different other effects cannot be ruled out, an internal origin 
seems to nicely account for the majority of observations \citep{kgm09}: 
this includes the shape of the light curve and superposed bumps 
(Fig. \ref{grb071031}),
the spectral hardening towards the optical wavelengths, 
the observed temporal decrease of the peak energy $E_{peak}$ 
between prompt emission and the flares, and the overall broadband flare 
spectrum from NIR to X-rays. 
The spectral similarities of the X-ray flares with the prompt phase 
suggest that they are later and softer examples of the prompt 
emission flares, and due to internal shocks. Thus, the simultaneous 
broad-band observation of GRB 071031 provides 
additional evidence that inner engine activity may last (or be revived)
over hours or days, at least for some bursts.

\begin{figure}[t]
\vspace{-0.3cm}
\hspace{-.8cm}\includegraphics[angle=270,width=10.0cm]{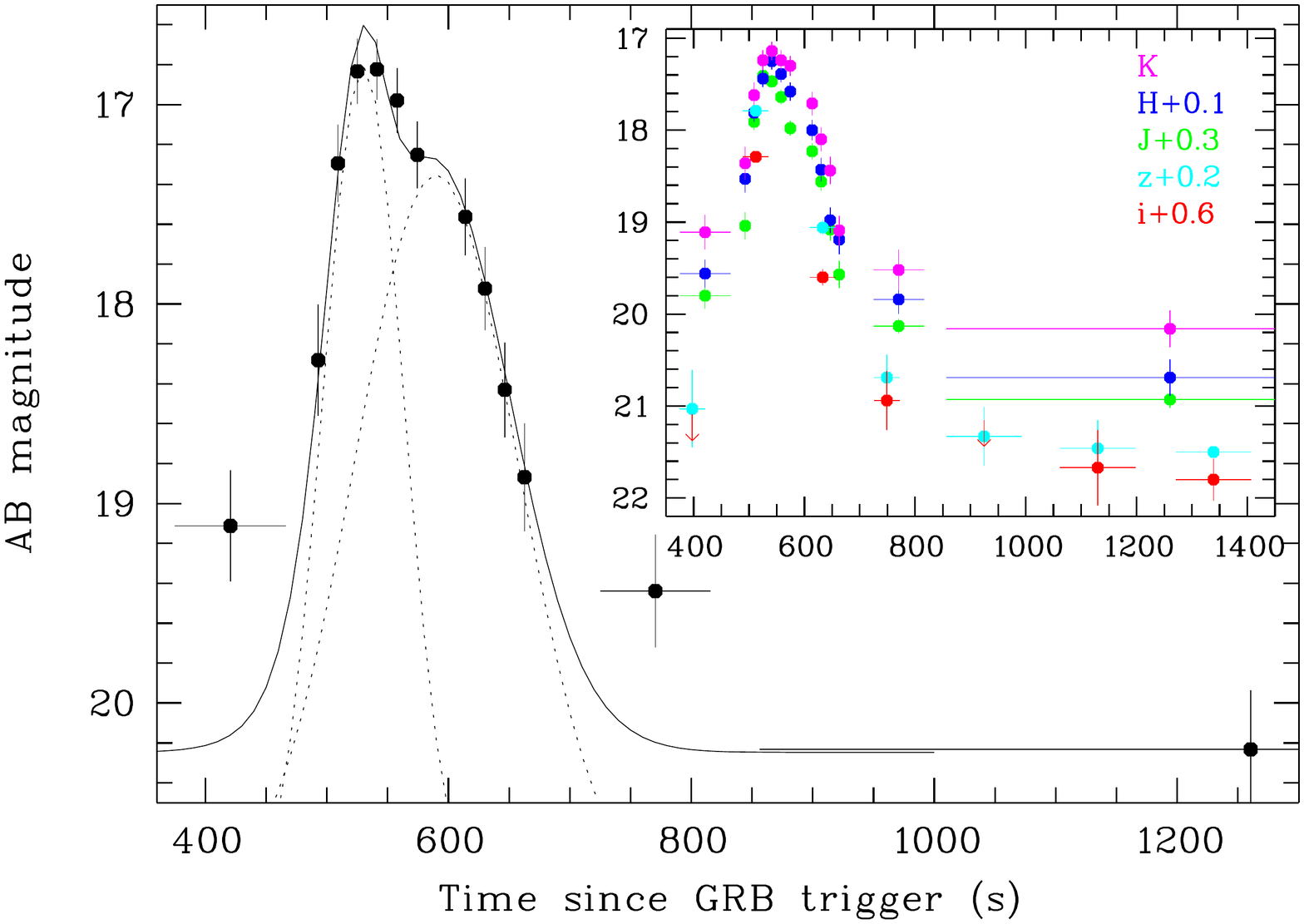}
\vspace{-.5cm}
\caption{Co-added GROND $JHK_s$ light curve (the inset shows the individual
light curves plus those of \ip\zp; \gp\rp\ are omitted due to Ly-absorption)
of the early flare in GRB 081029. During most of the flare, the individual
10 s integrations are shown. The model (solid line) consists of the 
sum of two Gaussians (dotted lines) with FWHM of 77 and 157\,s, respectively.
}
\label{grb080129}
\end{figure}

A much more extreme case was the early (first few hundred seconds)
optical/NIR emission after GRB 080129 \citep{gkm09}: prior to the
rising afterglow emission (peaking later than 7000 s after the GRB)
a strong and rapid flare was observed, with an amplitude of 3.5 mag,
and with a mean duration of 150 s. This was one of the rare occasions
where emission related to a GRB was bright enough in the NIR bands in 
each of the 10 s sub-integrations. This allowed us to resolve the flare 
into sub-structure (Fig. \ref{grb080129}), the shorter having a full-width
at half maximum (FWHM) of 77 sec. This is even more astonishing 
when considering the redshift of 4.3 for this GRB, i.e. the intrinsic
rest-frame FWHM was 15 sec. The simultaneous
observation in seven channels with GROND provides a SED from the optical 
to the near-infrared at a time resolution of once every minute.
The delay of the flare relative to the prompt GRB, its SED 
as well as the ratio of pulse widths suggest that it arises from residual 
collisions in GRB outflows \citep{vvm11}. 
Unfortunately, neither did {\it Swift}/XRT observe GRB 080129 at this time
(blocked by Earth), nor did we ever detect a similar flare in another
GRB in the following 8 years.

However, we did detect sudden intensity jumps in several GRBs at later times, 
between 10$^3$--10$^4$ s after the GRB, with amplitudes in the 1-3 mag range.
In these cases, the rise times were always much faster than the decay times.
The ``pulse'' shapes ranged from triangular to nearly rectangular,
thus justifying the term ``jump'' component, see e.g. GRBs 100621A 
\citep{gkn13}, 081029 \citep{Nardini+2011} or 100814A \citep{Nardini+2014}. 
With a similar SED and only little simultaneous emission 
at X-rays (consistent with the slope of the optical/NIR SED slope),
the same interpretation via residual collisions has been proposed.

\subsection{GRB jet structure and off-axis appearance}

Narrow jets, of order 5-20\degr\ opening angle, are usually
invoked for the interpretation of the observed GRB emission
primarily to reduce the otherwise huge inferred intrinsic energy
budget. The opening angle of these jets as well as their
radial energy distribution are then the next level of detail
which need to be determined in order to constrain the GRB energetics. 

It is usually assumed that all GRBs have the same universal
(with some dispersion of the parameters) structure, but that they
appear different because we see them under different observer angles $\theta_o$
\citep{Lipunov+2001}.
The jet structure is assumed to be axisymmetric, and is defined by the
radial distribution of the energy per jet unit solid angle $\epsilon(\theta)$,
and that of the Lorentz factor $\Gamma(\theta)$ of the emitting material.
In models for an inhomogeneous or a structured jet 
($\epsilon(\theta) \propto \theta^{-s}$, s$>$1), the initial bulk 
Lorentz factor,
the specific deceleration time and the radius are dependent on the 
distance from the symmetry axis of the jet \citep{Rhoads99, GranotKumar03}.
Thus, a geometric offset of the observers from the jet
symmetry axis has a distinct signature in the observed optical light curve.
Because of the relativistic beaming of the decelerating ejecta, an observer 
located off-axis 
will see a rising optical afterglow 
light curve at early times \citep[e.g.][]{Panaitescu1998, Granot2002}, with
the steepness of the rise being characteristic of the off-axis angle and the 
jet structure.
For the interpretation of afterglow light curves the structured jet is
often approximated by a two-component jet, i.e. narrow jet with high $\Gamma$
surrounded by a wider cone with small $\Gamma$.

\begin{figure}[t]
\includegraphics[width=8.6cm]{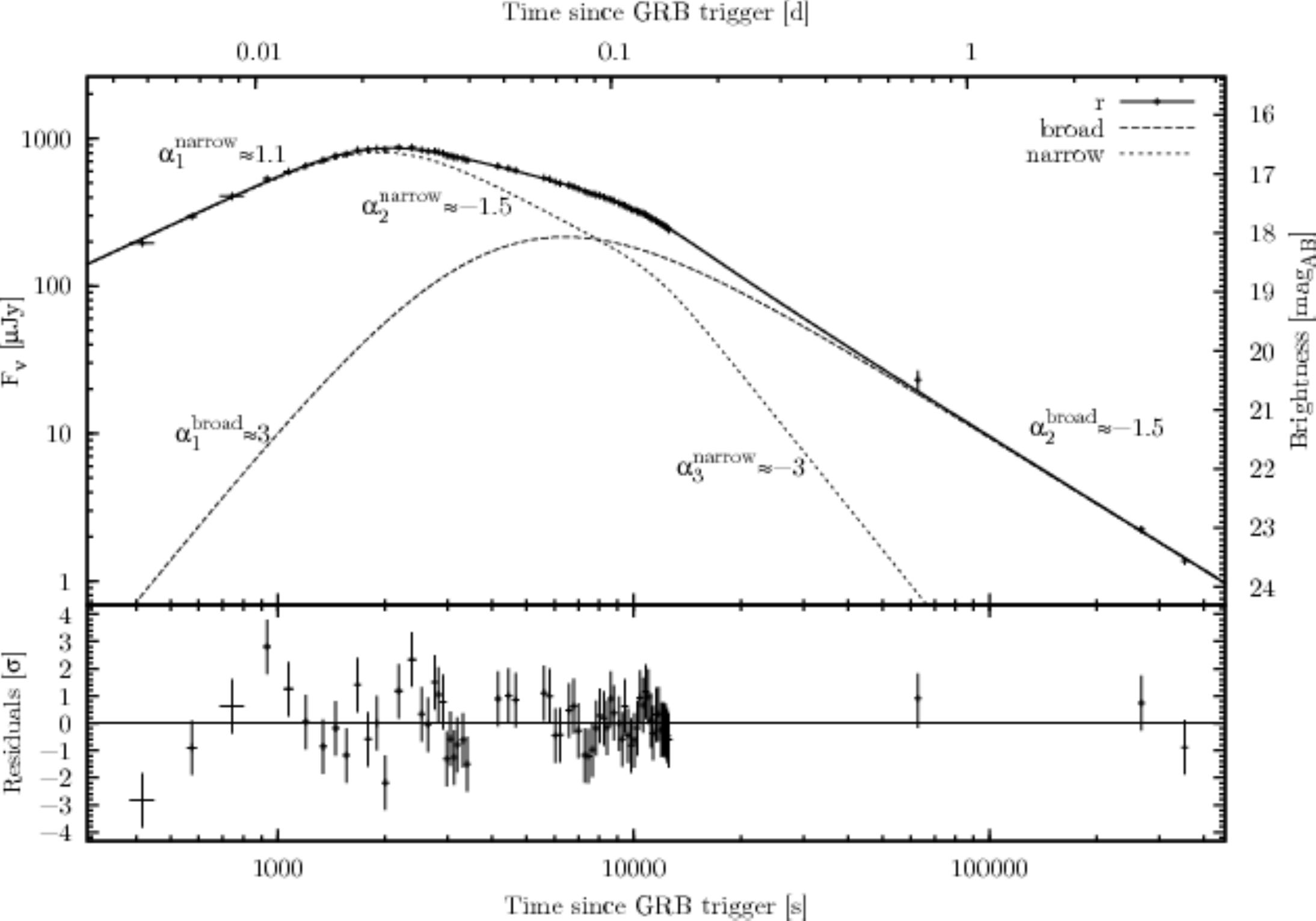}
\caption{GROND \rp\ light curve of the afterglow of GRB 080710 (upper panel)
with the best fit (solid line) of a two-component jet model with 
$\nu_m < \nu_{opt} < \nu_X < \nu_c$ for both components. 
The narrow jet (dotted line) is seen off-axis and produces a shallow rise
as its emission spreads during deceleration. The broad jet (dashed line)
is viewed close to on-axis with initial Lorentz factor $\sim$50 and
opening angle $>$10\degr, and has the expected steep rise during its
pre-deceleration phase. [From \cite{kga09}].
}
\label{grb080710}
\end{figure}

For two bright GRB afterglows, GROND data argue for a 
two-component jet structure as preferred interpretation.
The multi-band afterglow light curve of GRB 080710 (Fig. \ref{grb080710})
shows two salient features,
both achromatic to high precision: an early rise in its brightness, too shallow
to be caused by a jet in the pre-deceleration phase, and a turnover from a 
shallow to a steep decline without a change in spectral slope, thus 
incompatible with a jet break \citep{kga09}. The most natural explanation 
 is a two-component jet (Fig. \ref{grb080710}), with the
narrow component (2\degr-4\degr\ opening angle) viewed slightly off-axis, 
and the wider component with lower Lorentz factor dominating the late emission.
In this interpretation, the shallow decay phase is the result of the 
superposition of the narrow-jet afterglow and the rise of the broad jet
in its pre-deceleration phase \citep{kga09}.

GRB 080413B is well fit with an on-axis two-component jet model. The
narrow ultra-relativistic jet is responsible for the initial decay, and
the rising of the moderately relativistic wider jet causes a re-brightening
and dominates the late evolution of the afterglow \citep{Filgas+2011}.
The deduced jet opening angles are 2\degr\ and 9\degr\, respectively,
for the narrow and wide jet, and the initial Lorentz factors
$>$190 and 19. This model also explains the relative fluxes and spectral
shapes of the X-ray vs. optical/NIR emission as well as the chromatic
re-brightening due to the different spectral regime of the wide jet.

In both cases, the early and very accurate multi-colour light curves provided 
by GROND were essential in excluding alternative explanations, such as a 
reverse shock, emission during the pre-deceleration phase,
refreshed shock emission, or an inhomogeneous ISM density profile 
\citep{kga09, Filgas+2011}.

\subsection{Short-duration GRBs}

Since the afterglows of short-duration GRBs are substantially less
luminous, their discovery was accomplished only in 2005 \citep{Fox+2005}
with the advent of the fast and accurate localization with the Gehrels 
{\it Swift} Observatory \citep{Gehrels+2004}.
The faint optical afterglows also meant that small robotic telescopes
had little success and impact. Even with GROND at a 
2\,m class telescope, the detection rate of short GRBs is a factor 
of two smaller than that for long GRBs (see Fig. \ref{grondGRBs}),
and the detections typically do not extend beyond 2--3 days after 
the GRB. Yet, this was long enough for GROND to establish the first cases 
of clear jet-breaks in the afterglows of short-duration GRBs 
\citep{NicuesaGuelbenzu+2011, NicuesaGuelbenzu+2012}.
This provided first observational
hints that the jet opening angles in short GRBs are wider than those
in long-duration GRBs, as earlier suggested on theoretical grounds 
\citep{Aloy+2005}.

\begin{figure}[t]
\hspace{-0.5cm}\includegraphics[width=10.2cm]{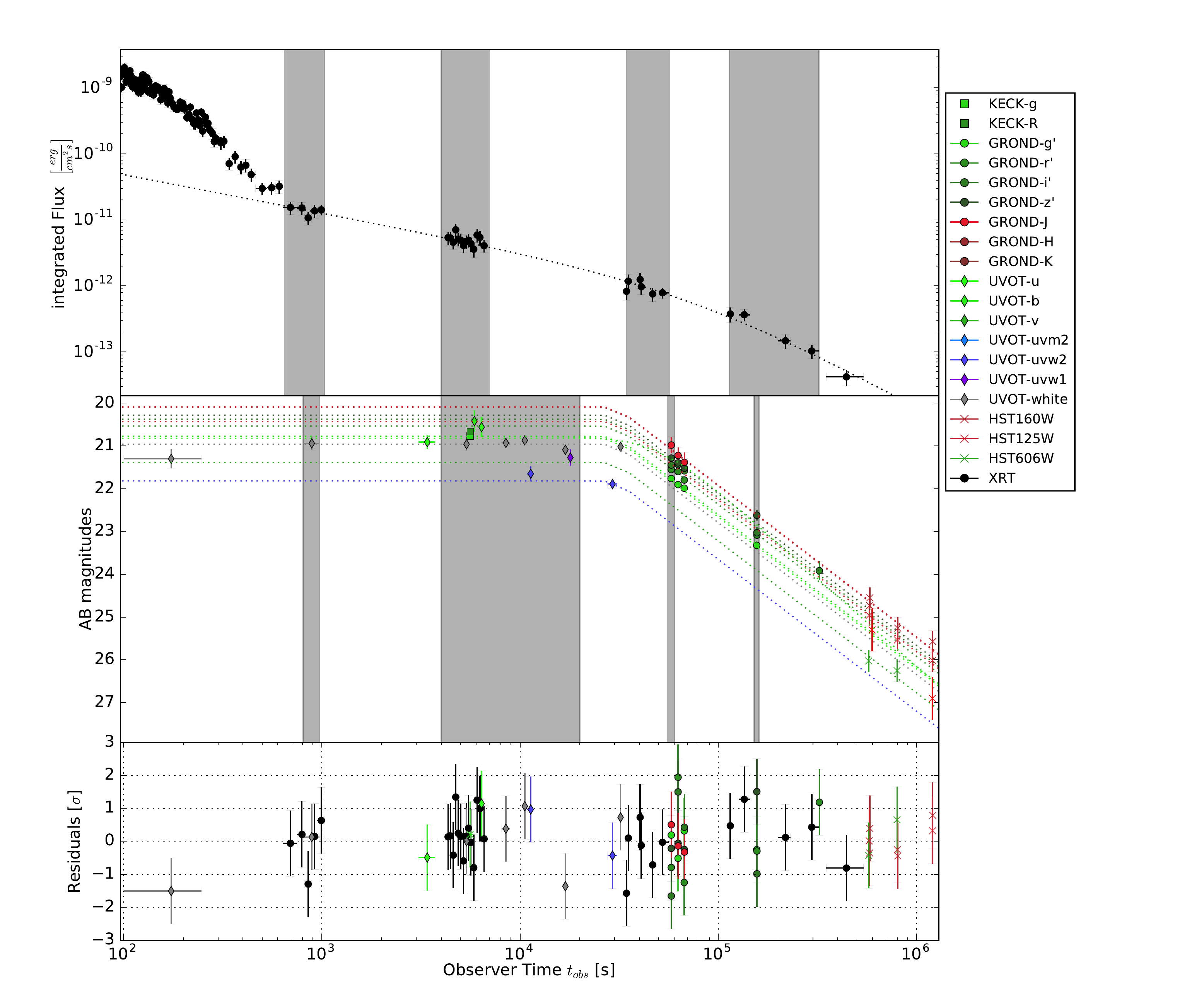}
\caption{X-ray and optical light curve of the
afterglow of the short-duration GRB 150424A.
The grey-shaded areas indicate the time slices used for the SED analysis
which together with the best-fit temporal slopes allows us to reject
all standard scenarios.
[From \cite{Knust+2017}].
}
\label{grb150424A}
\end{figure}

Apart from their duration and peak energy, short GRBs show many
phenomenological properties similar to long GRBs. Among those properties
is optical plateau emission, e.g. GRB 060313A \citep{Roming+2006}, 
GRB 061201A \citep{Stratta+2007}, or GRB 130603B \citep{Fan+2013}.
A particularly well-sampled example is GRB 150424A (Fig. \ref{grb150424A}),
where our GROND data provide convincing evidence for 
a uniform, nonspreading jet expanding into an ISM medium as a
self-consistent explanation \citep{Knust+2017}, where the jet is 
re-powered for 10$^4$ s with additional constant energy injection.
Within a factor of two,
this unique and very-long-duration energy injection in GRB 150324A
provides a similar energy input as the prompt GRB emission \citep{Knust+2017}.

\subsection{GRB hyper- and kilo-novae}

The association of hydrogen- and helium-free core-collapse supernovae (type Ic)
to long-duration GRBs, first seen in GRB 980425 / SN 1998bw \citep{Galama+1998}
and then conclusively observed for GRB 030329 / SN 2003dh 
\citep{Hjorth03, Stanek03} has established their relation to exploding 
massive stars. The kinetic energy of both supernovae was in excess of 
10$^{52}$ erg, a factor 10 larger than canonical SNIc, and thus 
earning the name 'hypernovae' (though that name has been in use already 
since the early 80ies). The inferred rates of SNIc and GRBs differ by at
least a factor of 100 (depending on the actual GRB beaming angle), and
the still debated question is what causes a small fraction of supernovae
to produce a GRB \citep{Woosley+1999}?

Despite having now over 700 GRBs with an observed optical afterglow, less 
than 50
GRB-supernovae are known to varying degree of confidence, and only 11
of these have strong spectroscopic evidence in the optical \citep{Cano+2017}. 
Partially, this is due to the fact that the supernova light is getting difficult
to observe beyond redshift $z \sim 0.5$, but certainly also the
lack of systematic late-time ($\sim$10 days) optical monitoring 
implies that many GRB-supernovae are missed. 
With our concept of following each observable GRB as long as 'something'
is detected, GROND observations at least doubled the annual rate of
discovered GRB-supernovae 
\citep[e.g.][]{Cano+2014, Olivares+2012, Olivares+2015, Klose+2018}.

\begin{figure}[t]
\vspace{-2.cm}\hspace{-0.7cm}\includegraphics[width=10.2cm]{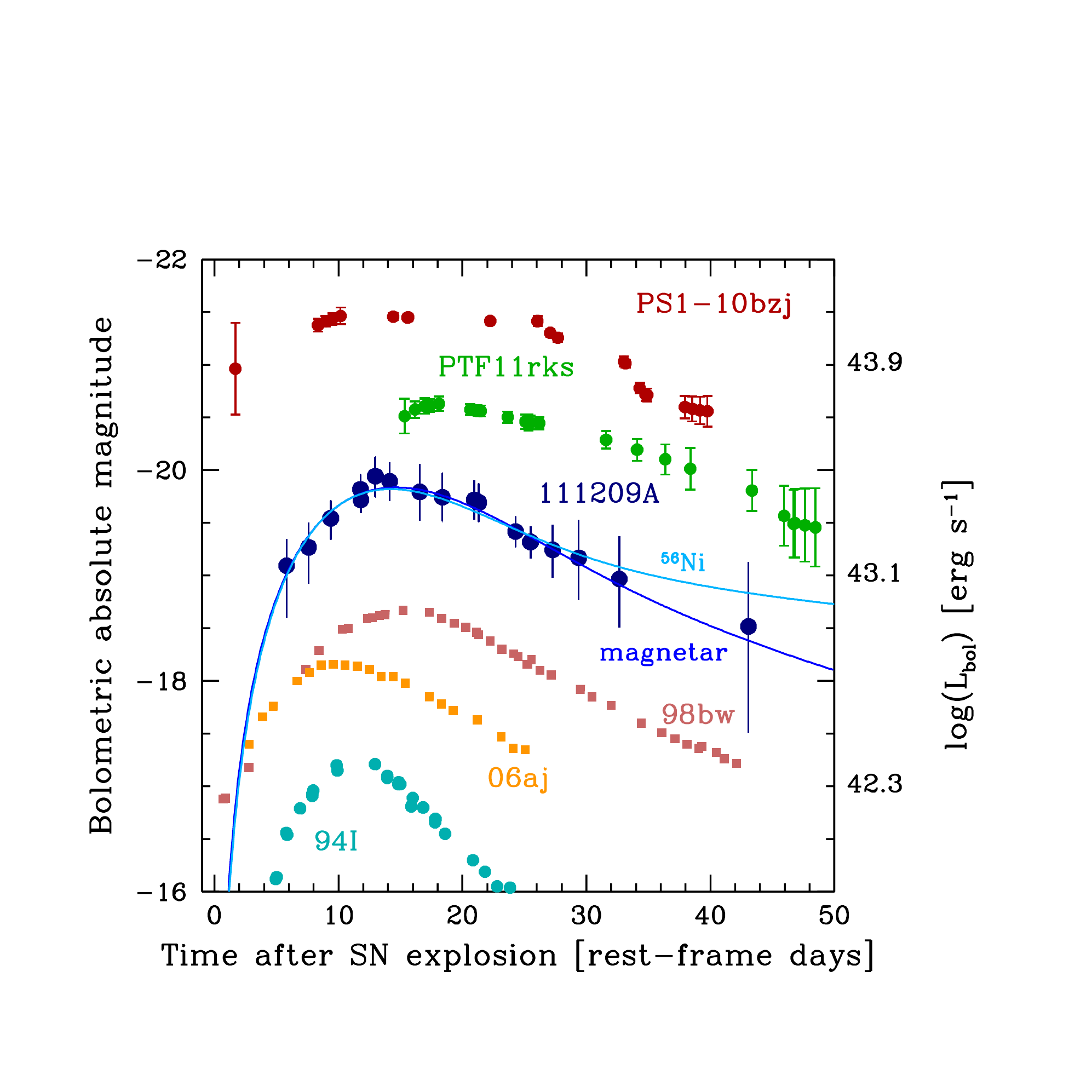}
\vspace{-1.5cm}
\caption{Afterglow- and host-subtracted bolometric light curve of the supernova 
related to GRB 111209A as observed with GROND (\gp\rp\ip\zp$J$) in the 
2300--8000 \AA\ rest frame, compared with bolometric light curves of
GRB 980425 / SN 1998bw \citep{Galama+1998},
XRF 060218 / SN 2006aj \citep{Pian+06},
the standard type Ic SN 1994I \citep{Sauer06},
and the superluminous supernovae PTF11rks \citep{Inserra13} 
and PS1-10bzj \citep{Lunnan13}.
The dark blue line shows the best-fitting synthetic light curve computed 
with a magnetar injection model based on \cite{KasBil10}.
The bright blue line shows the best-fit $^{56}$Ni light curve.
[From \cite{Greiner+2015}].
}
\label{SN2011kl_lc}
\end{figure}

GRB 111209A / SN 2011kl turned out to be particularly interesting,
as the prompt emission had an ultra-long duration ($>$4 hrs), revealed
with the Konus detector on the WIND spacecraft \citep{Golen11}.
The GRB occurred at a redshift of $z=0.677$
\citep{lts14}, as determined from afterglow spectroscopy.
Its integrated equivalent isotropic energy output is 
$(5.7 \pm 0.7)\times10^{53}$\,erg \citep{Golen11},
corresponding to the bright end of the distribution of long-duration GRBs.
Several models had been proposed to explain the ultra-long duration of 
GRB 111209A (and a few others), but the otherwise inconspicuous spectral and 
timing properties of both, the prompt and afterglow emission as well as
the GRB host galaxy properties, provided no obvious clues to distinguish 
among these \citep{gsa13, lts14, nks13}.

The corresponding GRB-supernova SN 2011kl was a factor of $>$3x more 
luminous and its spectrum
distinctly different from other type Ic supernovae associated with 
long-duration GRBs. The slope of the optical continuum resembles those of 
super-luminous supernovae, but the light curve evolved much faster 
(Fig. \ref{SN2011kl_lc}). 
The combination of high bolometric luminosity but low
metal-line opacity cannot be reconciled with typical SN Ic, like in
all previous GRB-SNe. Instead, it can be reproduced by invoking a magnetar, 
a strongly magnetized neutron star,
which injects extra energy \citep{Greiner+2015, Kann+2016}.
The detection of a supernova associated with the ultra-long GRB 111209A
immediately rules out a tidal disruption event as the
origin of GRB~111209A \citep{lts14}.
Also, blue supergiants \citep{nks13} are ruled out as progenitors,
since they show hydrogen in their spectra
and have substantially different light curves \citep{Kleiser11}, 
inconsistent with our observations.
Instead, GRB 111209A / SN 2011kl provides a link between GRB/SNe on the one 
hand, and ultra-long GRBs and superluminous SNe on the other.

\begin{figure}[t]
\vspace{-0.2cm}
\hspace{-1.1cm}\includegraphics[angle=270, width=10.6cm]{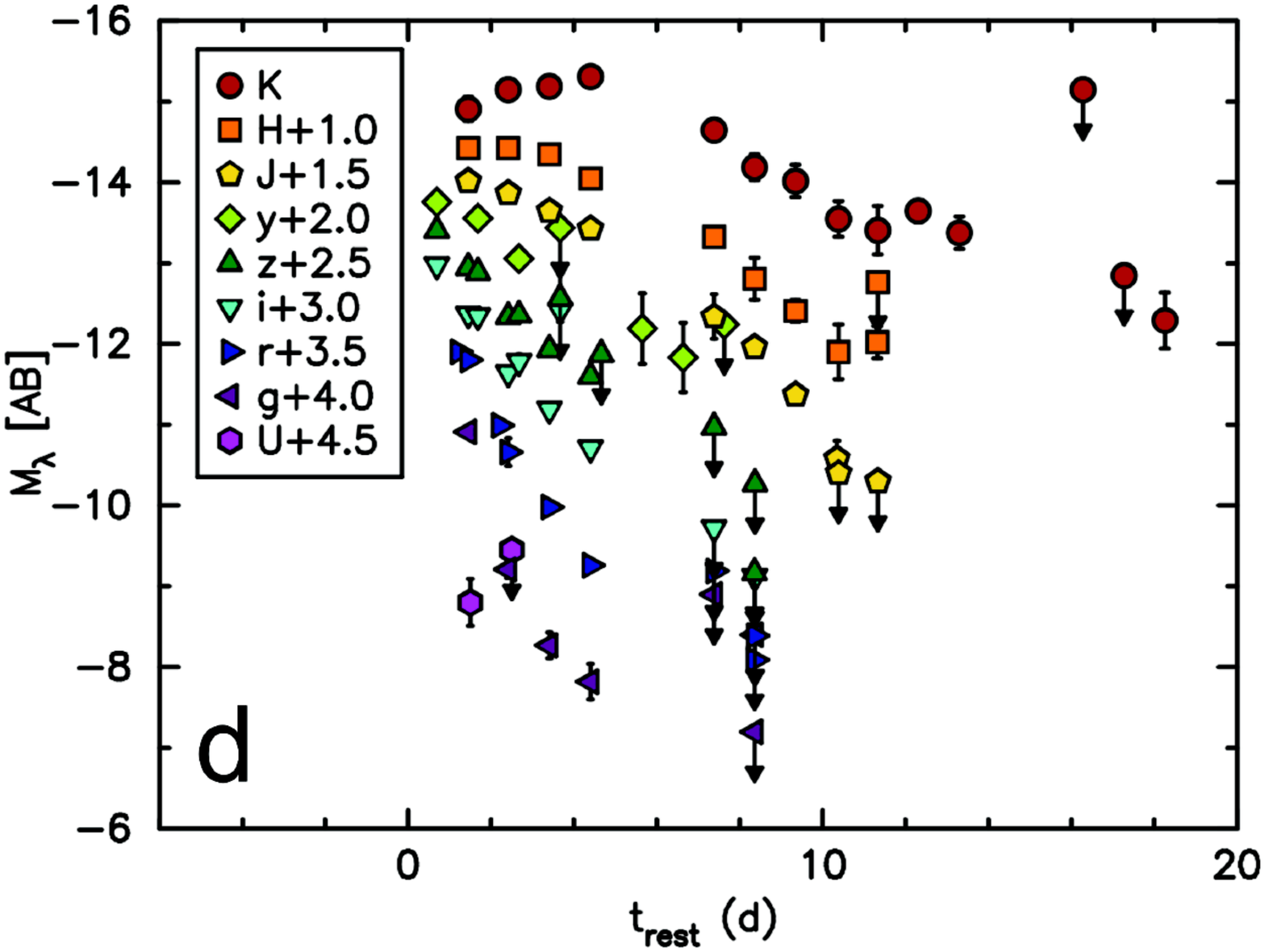}
\vspace{-0.5cm}
\caption{GROND lightcurve of the kilonova of GRB 170817A,
with selected $U-$, and $K-$band (NTT) as well as $y-$band (PS1) data points,
corrected for Galactic foreground extinction and
transformed to absolute AB magnitudes using the distance of 40 Mpc.
[From \cite{Smartt+2017}].} 
\label{170817_lc}
\end{figure}

The short-duration burst GRB 170817A became famous for its gravitational wave
detection \citep{LIGO+2017}. It occurred extraordinarily nearby at 40 Mpc,
or $z=0.009$ \citep{LIGO_APJL_MM, Burgess+2017}, thus 
its kilonova was in reach for even small telescopes. Yet,
the short visibility period per night ($\sim 1$ hr) made GROND's
simultaneous 7-channels again particularly useful (Fig. \ref{170817_lc}). 
As the kilonova faded, the SED rapidly changed from blue to red, 
and a higher-opacity, lanthanide-rich ejecta may have contributed 
to the late-time emission. 
The decline is measured to have a power-law slope of 1.2$\pm$0.3 
\citep{Smartt+2017} which is 
consistent with radioactive powering from $r$-process nuclides. 
The derived physical parameters broadly match the theoretical predictions of 
kilonovae from neutron-star mergers \citep[e.g.][]{Metzger+2010, Kasen+2013}.

\section{Identification of transient or steady high-energy sources}

\subsection{X-/$\gamma$-ray Transients}

Most objects which presently trigger high-energy missions with their
X-ray or  $\gamma$-ray transient behavior 
\citep{Kennea2015, Abdollahi+2017, Negoro2017}
are accreting systems,
exhibiting either thermal emission from accretion disks, bremsstrahlung
due to shocks in winds, or synchrotron emission in jets.
Consequently, enhanced emission at other wavelengths is accompanying
the X-/$\gamma$-ray transients, often at optical and/or near-infrared
wavelengths, originating predominantly either in the accretion disk or the jet. 
Galactic sources do occur mostly in the disk of
the Milky Way, making near-infrared observations more promising due to 
the smaller affect of absorption by dust.

With varying degree of effort over the years, newly discovered X-ray transients
from {\it Swift}, MAXI or the XMM-{\it Newton} slew survey were followed up
with GROND. Typically, results were published within hours (e.g. 32 
Astronomical Telegrams between 2008-2016), with about 50\% of these
reporting discoveries of the optical/NIR counterpart.

\begin{figure}[b]
\includegraphics[width=8.5cm]{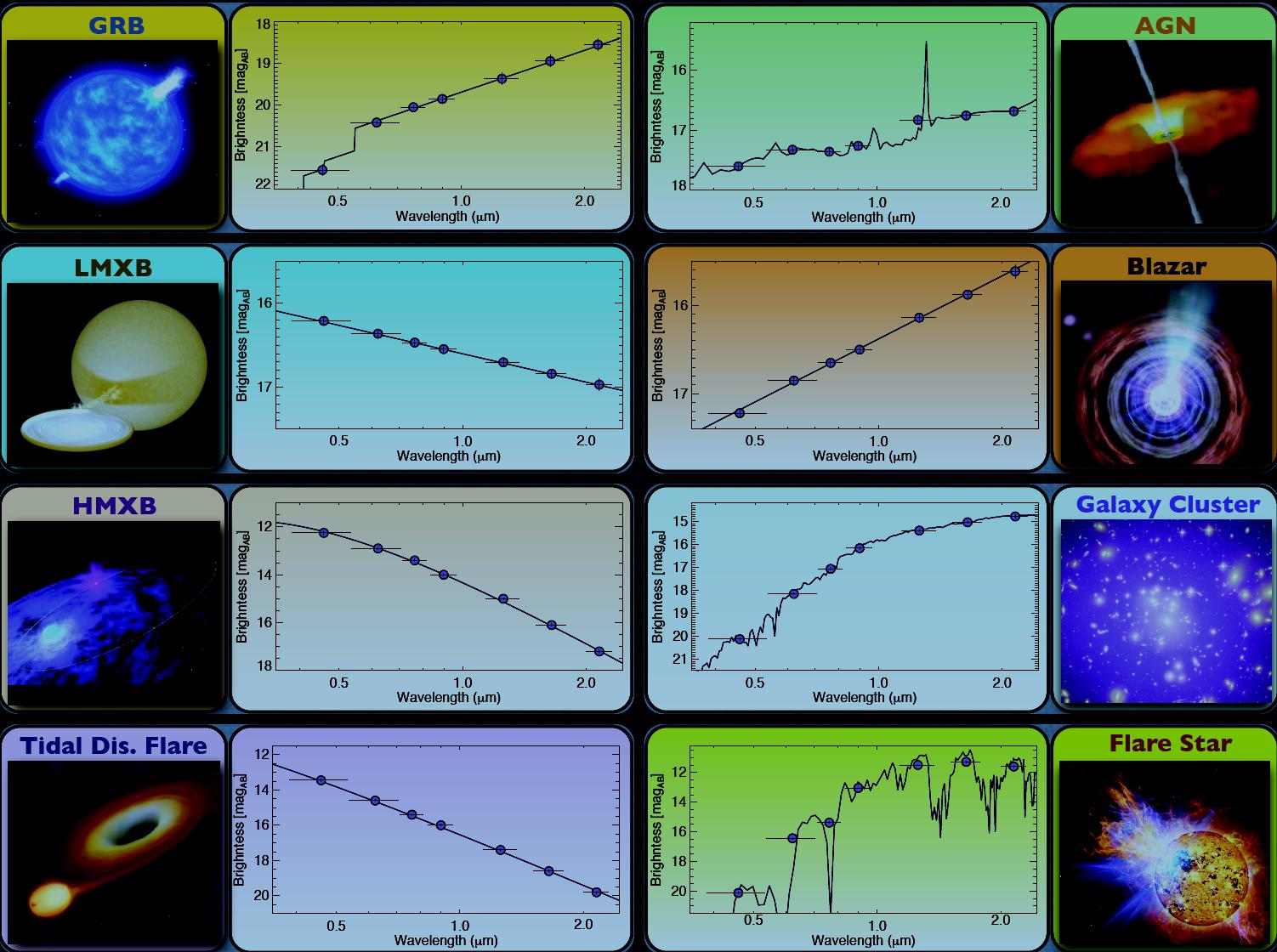}
\caption{Source typing power of GROND via its simultaneous 7-channel
imaging capability [From \cite{Rau2012}].
}
\label{SED-typing}
\end{figure}

These identifications predominantly rely on the detection of a
new or substantially brightened source relative to a reference catalog. 
The 7 simultaneous GROND channels and the corresponding spectral energy
distribution are particularly useful in assigning source classes
even for non-variable objects. As Fig. \ref{SED-typing} shows,
accreting sources (blue SEDs) can be easily distinguished from
blazars or GRB afterglows (red SEDs), or stellar objects.
One example was INTEGRAL trigger 5994, which reported a discovery
of a long-duration GRB \citep{Mereghetti+2010}, but was shown
to be a variable object with a stellar SED; this allowed us to reject 
the GRB classification \citep{Updike+2010}.

\subsection{Tidal disruption events}

Tidal disruption events (TDEs) are obviously another very intriguing type of
transient which were followed-up with GROND at various occasions
\citep[e.g.][]{Cappelluti+2009, Komossa+2009, Merloni+2015}.
TDEs ensue when a star gets close to a supermassive black hole (SMBH) in the 
center of a galaxy, and is ripped apart by the tidal forces of the black hole.

TDEs promise to help solving several astrophysical questions,
among others about the accretion formation and physics in early stages
of TDE evolution \citep{Lodato+2015}, 
the formation and ejection of relativistic jets,
the prevalence of (dormant) single or binary SMBHs in 
galaxies \citep{Komossa2017},
or verifying signatures of General Relativity (delay in accretion disk
formation or quasi-periodic modulations at X-rays; \citealt{Stone+2018}).

Observations of Swift J2058.4+0516, the possible second relativistic 
TDE discovered, revealed faint optical emission despite small intrinsic
extinction, suggesting that either the outflows are extremely narrowly
collimated, or that only a small fraction of tidal disruptions generate
relativistic ejecta \citep{Cenko+2012}.

The unusual transient OGLE16aaa, recently detected by the Optical 
Gravitational Lensing Experiment (OGLE-IV) survey \citep{Wyrzykowski+2016,
Greiner+2016},
shows many optical features similar to other TDEs. 
The spectral properties and photometric history of the host galaxy suggest 
that OGLE16aaa belongs to a sub-class of TDEs which is
associated to weakly or only recently active SMBHs. 
This class might provide a connection between TDEs from quiescent SMBHs 
and flares observed as `changing-look quasars', if the latter 
are interpreted as TDEs. In this case, the previously applied selection
criterion for identifying a flare as a TDE to have come from an 
inactive nucleus, 
would represent an observational bias, thus affecting 
TDE-rate estimates \citep{Wyrzykowski+2017}.

\subsection{Steady sources}

The characterization and typography of sources via the 7-channel SEDs can 
obviously be used  also
for steady sources, i.e. for the optical/NIR identification of X-/$\gamma$-ray
sources. Applications of this possibility have not yet been published, but
observations have been taken for unidentified ROSAT sources; 
a wider use is anticipated for new X-ray sources which the upcoming
eROSITA survey will discover.

Identifying individual galaxies as belonging to the same galaxy cluster by 
color 
selection is a wide-spread method. Using GROND with its seven simultaneous 
channels allows not only for a substantially more secure selection (by using
the full SED instead of the usual two filters), but also a photometric
redshift estimate of individual cluster members (see e.g. 
\citealt{Pierini+2012} for a cluster at z=1.1). 

A more unusual application of this method was the search for a
suspected companion star of a 
neutron star formed in a supernova which created the remnant RCW 86.
GROND observations identified such a candidate with uncommon SED-shape
which allowed to justify 
follow-up VLT spectroscopy. This in turn revealed that this neutron star 
companion was strongly polluted with calcium and other elements.
Combining all constraints suggests that the progenitor of the supernova that 
produced RCW 86 was a moving star which exploded near the edge 
of its wind bubble and lost most of its initial mass because of 
common-envelope evolution shortly before core collapse 
\citep{Gvaramadze+2017}.

\section{X/$\gamma$-ray binaries}

\subsection{Heating in a $\gamma$-ray pulsar}

Color variations over orbital phase in X-ray or $\gamma$-ray binaries
are frequently observed. Simultaneous multi-color observations such as with
GROND are particularly rewarding for short orbital periods. One interesting
example is the $\gamma$-ray black widow pulsar PSR J1311-3430
\citep{Romani+2012}. 
Black widow pulsars are binaries consisting of a millisecond pulsar and a very 
low-mass star (brown dwarf), in which the strong radiation from the
neutron star ablates the companion, thus leading to outflows strong enough 
to eclipse the pulsar signal for a good fraction of the orbit.
Black widow systems allow an accurate neutron star mass determination,
and consequently constraints on the equation of state of neutron stars
\citep{Lattimer+2011}.

\begin{figure}[t]
\includegraphics[width=8.4cm]{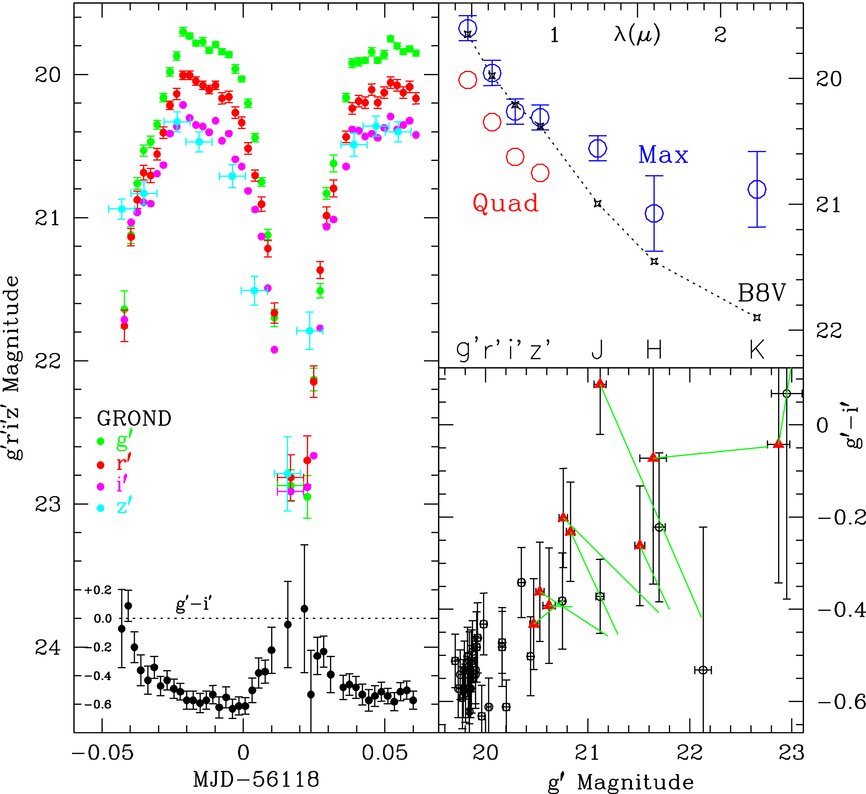}
\caption{GROND \gp\rp\ip\zp\ light curve (left) with the \gp-\ip\ color
change at the bottom, showing reddening during the minimum, the pulsar
superior conjunction. The upper right panel shows the large NIR excess of the 
SED in comparison to a B8V star. The lower right shows the color variations
with brightness, where green lines and red triangles mark epochs with
$\Delta$\gp$>$0.2 mag over the quiescent magnitude.
From \cite{Romani+2012}. \textcopyright AAS. Reproduced with permission.
}
\label{widow}
\end{figure}

The bright $\gamma$-ray source 2FGL J1311.7-3429 (3FGL J1311.8-3430), 
known since the early EGRET mission (3EG J1314-3431), was identified as 
a 2.5 ms pulsar in Fermi-LAT data once an optical counterpart had been found.
With an orbital period of 94 min, it shows more than 3 mag amplitude
variations. Spectroscopy revealed that the companion is a bloated, 
Roche-lobe filling substellar object with a He-dominated photosphere,
while no hydrogen is seen. GROND photometry showed a strong color variation
with orbital period (Fig. \ref{widow}), and a reddening at the pulsar's 
superior conjunction 
by \gp-\rp\ $\sim$0.6 mag. At maximum light, the colors in the visible 
wavelength range are comparable to those of a B8 star, while in the NIR
there is a large excess. This suggests a large emitting area at low 
temperature. One possible source is the evaporative wind, reprocessing
the pulsar power into the optical/NIR. Also, short-term variability
(flares) are very red, suggesting a variable wind off the companion.

The flat part of maximum light is a challenge to light curve fits:
neither cold nor hot asymmetric spots helped to improve the fits,
suggesting additional physics to be at play. As these additional
model components also affect the best-fit inclination, the neutron
star mass estimate remains rather poor: 
depending on the exact modelling of the near-infrared flux,
we obtained 2.2-2.8 \msun\ \citep{Romani+2012}.
Later, more detailed phase-resolved spectroscopy and more sophisticated
light curve models also allow a mass as low as 1.8 \msun\ 
\citep{Romani+2015}. Yet, the neutron star mass determined for this black widow
remains interestingly high.

The {\it Fermi} satellite has proven to be efficient in finding
black widows with short orbital periods, with another 100 candidates
waiting for careful analysis. In an exploratory search for another
dozen candidates, only one more clear example could be identified via
few-hour GROND light curves  \citep[e.g.][]{Salvetti+2015, Salvetti+2017}.

\subsection{The closest jet source?}

During the search for the counterpart of a serendipitous Chandra X-ray 
source with an X-ray jet (CXO J172337.5-373442), 
a candidate in the optical/NIR was identified 
with GROND observations. Consistent values of visual extinction (as determined
from the GROND SED) and hydrogen column density (as determined from the
X-ray spectrum) as well as the spatial coincidence suggest that the
optical source is associated with the X-ray source. The good match of an
extrapolation of the GROND NIR fluxes to that of a nearby Spitzer source 
suggests
an association as well, with the full SED being consistent with a G9 V
star at a distance of 330$\pm$60 pc. Since the observed X-ray luminosity cannot
be explained in terms of emission from a single G9 star, it is likely that
CXO J172337.5-373442 is an accreting compact object in a binary system
\citep{Mookerjea+2010}. This makes it the nearest known resolved
X-ray jet from a binary system which is not a symbiotic binary.
The implied very low X-ray luminosity of only 7$\times$10$^{30}$ erg s$^{-1}$
(assuming isotropic emission) is at odds with the standard concept
of jet ejection in 'high-states' of the accretion disk. Even if this
system is a cataclysmic variable, the jet was launched in a state
of quiescence.
This implies that such jets are more ubiquitous than previously thought,
because they are difficult to detect at much larger distances 
\citep{Mookerjea+2010}.

\begin{figure*}[t]
\includegraphics[width=11.cm]{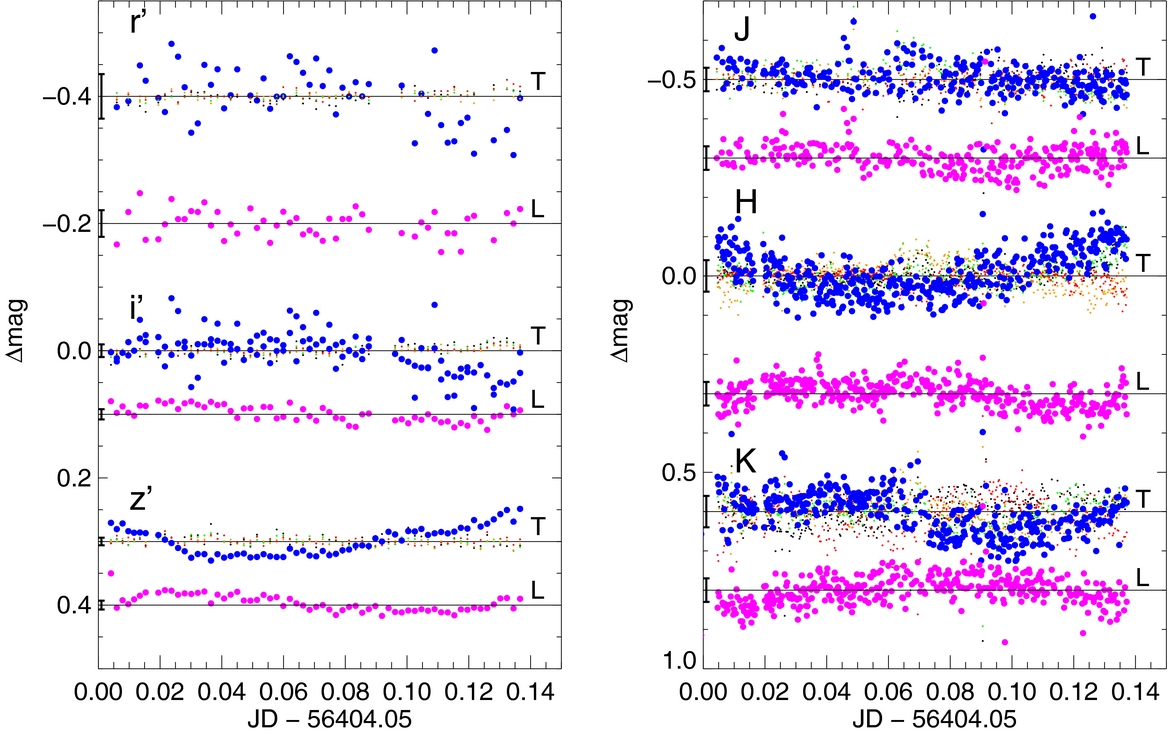}
\includegraphics[width=6.3cm]{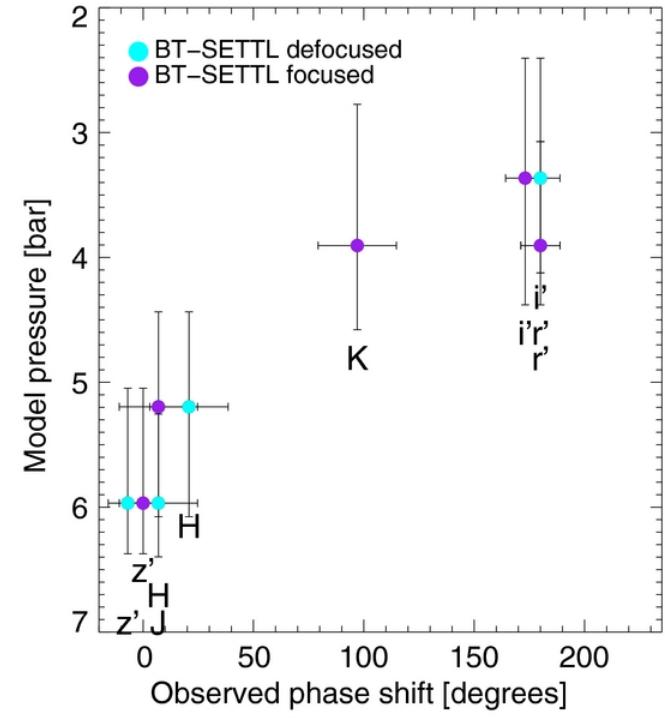}
\caption{{\bf Left and Middle:}
GROND light curves of the L7.5 (red) and T0.5 (blue) component,
respectively. Error bars are plotted at the very beginning of each light 
curve, and a (normalized) example residual light curve is shown in 
grey small dots.
{\bf Right:} The atmospheric pressure dependence on phase shift based
on the one-dimensional model of \cite{Allard+2012} 
[From \cite{Biller+2013}]. \textcopyright AAS. Reproduced with permission.
}
\label{browndwarflc}
\end{figure*}

\section{Brown dwarfs}

\subsection{Ross 458C: A benchmark T8-9 brown dwarf}

More than half of all stars (brown dwarfs included) have masses below 0.2 \msun.
The formation mechanism of these stars is uncertain, with theory suggesting
turbulent fragmentation, ejection of protostellar embryos, disc fragmentation 
or photo-erosion of prestellar cores \citep{Stamatellos2017}. Brown dwarfs
are objects below the hydrogen burning limit (mass range of 0.01--075 \msun).
The oldest brown dwarfs could be as old as the first generation of stars
that formed in the Universe. Brown dwarf studies have gained momentum with
the discovery of objects with decreasingly lower temperature and the
growing evidence that brown dwarfs and giant gas planets overlap in masses
and global temperature \citep{Chabrier+2014}.

Benchmark brown dwarfs are systems with well-known properties such as
effective temperature, parallax, age and metallicity. GROND follow-up of
candidates from a search of the DR5+ release of the UKIRT Deep Infrared 
Sky Survey revealed an object which shared its large proper motion with 
an active M0.5 binary at 102 arcsec distance, forming an hierarchical 
low-mass star and brown dwarf system \citep{Goldman+2010}. With a mass
of only 14 Jupiter masses and a distance of 11.4 pc, this young (less than
1 Gyr) system is a promising target to constrain the evolutionary
and atmospheric models of very low-mass brown dwarfs 
\citep{Goldman+2010, Burningham+2011}.

\subsection{Weather on the nearest brown dwarf}

Luhman16AB  or WISE J104915.57-531906.1AB is the closest (2.0$\pm$0.15 pc) 
brown dwarf pair
(1\farcs5 or 3 AU separation) with an L7.5 primary and a T0.5 secondary
\citep{Luhman2013}
and thus a prime target to search for dusty cloud structure break-up
\citep[][and references therein]{Biller+2013}.
Two sets of 4-hr observations each with GROND in April 2013, 
revealed anti-correlated variability
between different filters (Fig. \ref{browndwarflc}), 
as well as a phase offset of the $K$ band
light curve relative to $H$ and \zp\ \citep{Biller+2013}.
This offset is correlated with atmospheric pressure, as it can be probed in 
each filter band (right-most panel of Fig. \ref{browndwarflc}), as 
estimated from one-dimensional atmospheric models.
Follow-up CRIRES/VLT observations clearly show spectroscopic variability
over the rotation phase, and Doppler imaging makes this
patchy global cloud structure visible in the stellar surface map of
the B component of the system \citep{Crossfield+2014}.

\begin{figure*}[t]
\includegraphics[width=17.5cm]{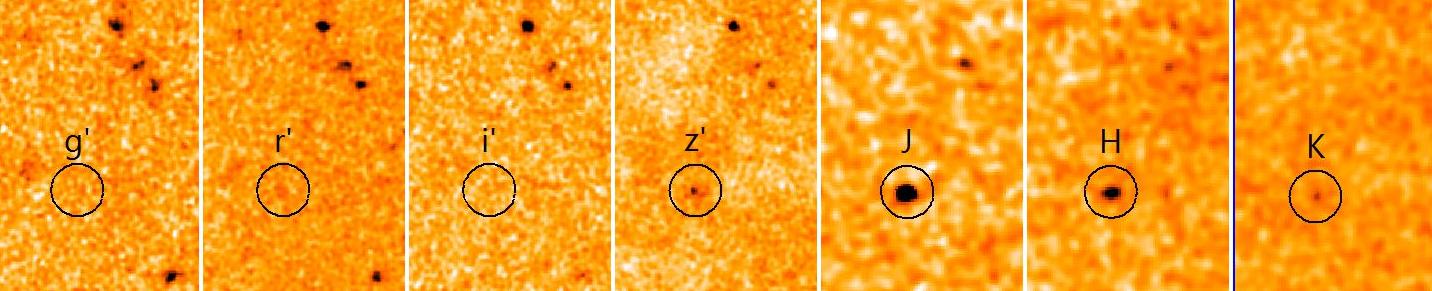}
\caption{GROND 7-channel finding chart of the brown dwarf J021003.48-042512.7;
North is at the top, and East to the left. The image sizes are 
23\asec$\times$33\asec.}
\label{BD_fc}
\end{figure*}

\subsection{New identification: J021003.48-042512.7}

\begin{figure}[b]
\hspace{-1.cm}\includegraphics[width=7.5cm,angle=270]{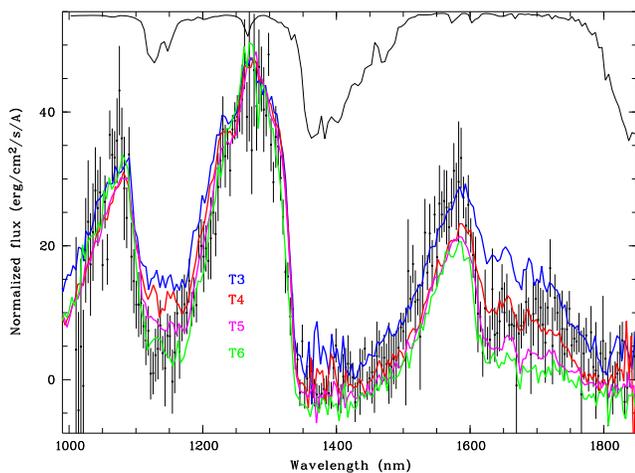}
\vspace{-1.0cm}
\caption{X-shooter spectrum of J021003.5-042512 with template
spectra of selected T dwarfs overplotted. 
\label{BD_XS}}
\end{figure}

The spectral energy distributions of brown dwarfs are similar to 
those of quasars at redshifts around 6, and thus are a frequent
contaminant of color-color search algorithms for high-z QSOs.
Since GRB afterglows were usually observed until they were not detectable
anymore in a 1-2 hr exposure, stacking of the individual exposures
give deep images around the GRBs observed since 2007.
During the search for high-z QSOs in these deep GRB fields, one particularly 
interesting example was  J021003.48-042512.7 ($\pm$0\farcs5) 
which has been found in a stack of 160 min GROND exposure
of GRB 131011A (Fig. \ref{BD_fc}). With AB magnitudes of \zp = 23.3$\pm$0.1,
$J$ = 20.9$\pm$0.1, $H$ = 21.1$\pm$0.1, $K$ = 21.2$\pm$0.3 mag,
it is close to T dwarfs or z$\sim$7 QSOs in color-color space.
We thus obtained a short (10 min exposure) VLT/X-shooter spectrum on 
13 Feb. 2014 which clearly solved the ambiguity (Fig. \ref{BD_XS}): 
a NIR spectral type of T5$\pm$1 provides the best match for this brown dwarf
using the templates of \cite{Burgasser+2006}.
Using $M_{\rm H} = 14.8$ mag (Vega) from \cite{Kirkpatrick+2011},
the distance of the brown dwarf J021003.48-042512.7 is 100$\pm$30 pc.
Being about 3 mag fainter than the WISE limit, this is one of the
most distant T dwarfs known.

\section{Exoplanets}

More than 20 years after the discovery of the first planet outside our solar 
system, more than 2000 exoplanets with very diverse properties have been 
discovered. With this large population, the field has moved from discovery
to characterization. Obviously, the largest interest lies in the study
of the atmospheric conditions, including their temperatures, albedos, 
compositions and cloud structures. But for the basic geometric properties 
like orbital periods and masses, transit measurements are one of the most 
important methods \citep{Cameron2016}.

\begin{figure}[b]
\includegraphics[width=8.5cm]{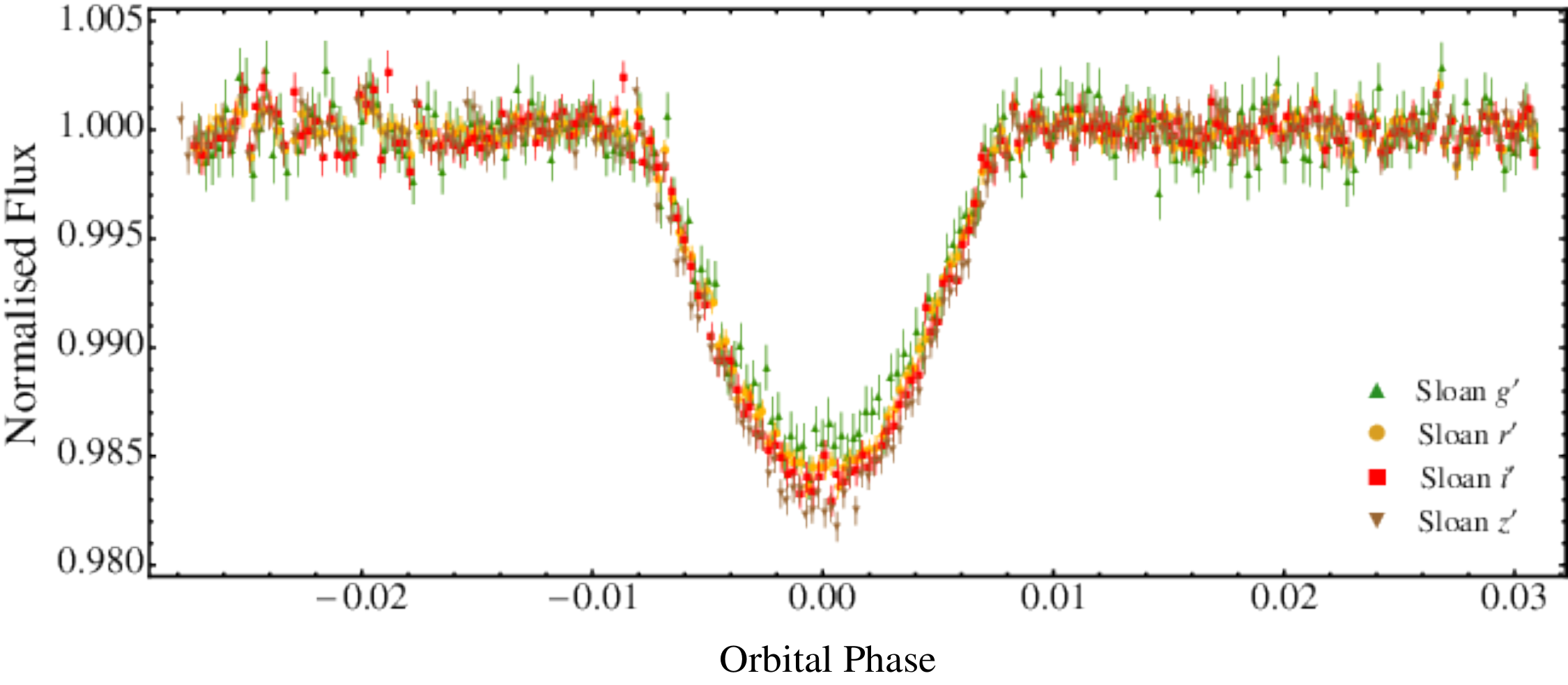}
\caption{GROND \gp\rp\ip\zp-band light curves of one WASP-67b eclipse, 
showing its wavelength-dependence: the bluer the color, the shallower 
the transit, as expected for a grazing eclipse, as limb darkening 
is stronger at bluer wavelengths. [Adapted from \cite{Mancini+2014}].
}
\label{WASP-67}
\end{figure}

Exoplanet transit observations benefit from simultaneous multi-filter imaging
in several ways. 
Firstly, it safely distinguishes proper transits from potential
blends between a star with a faint eclipsing-binary system.
This was nicely demonstrated by \cite{Snellen+2009} for OGLE2-TR-L9
where the mother star turned out to be an early F-star \citep{Lendl+2010}.
Secondly, it allows to recognize flares or spots on the mother star
which otherwise affect the interpretation of the light curve 
\citep[e.g.][]{Mancini+2013, Mohler-Fischer+2013}.
Thirdly, it provides evidence for grazing eclipses, since the limb
darkening predominantly affects the bluer wavelengths 
\citep[see Fig. \ref{WASP-67};][]{Mancini+2014}.
Furthermore, GROND's coverage of the NIR wavelengths enables the measurement
of the vertical temperature profile via flux ratios to the mother star, 
since layers at different depth are simultaneously probed at different 
filter bands \citep{Chen+2014}. 
Last but not least, differences in the ingress and egress slopes can 
be used to infer basic chemical ingredients of the planetary atmosphere.
Overall, exoplanet studies are likely the science topic with the largest
use of GROND observing time over the last decade.

\section{Blazars: Photometric redshifts down to $z \sim 1$}

\begin{figure}[b]
\hspace{-0.3cm}
\includegraphics[width=9.cm]{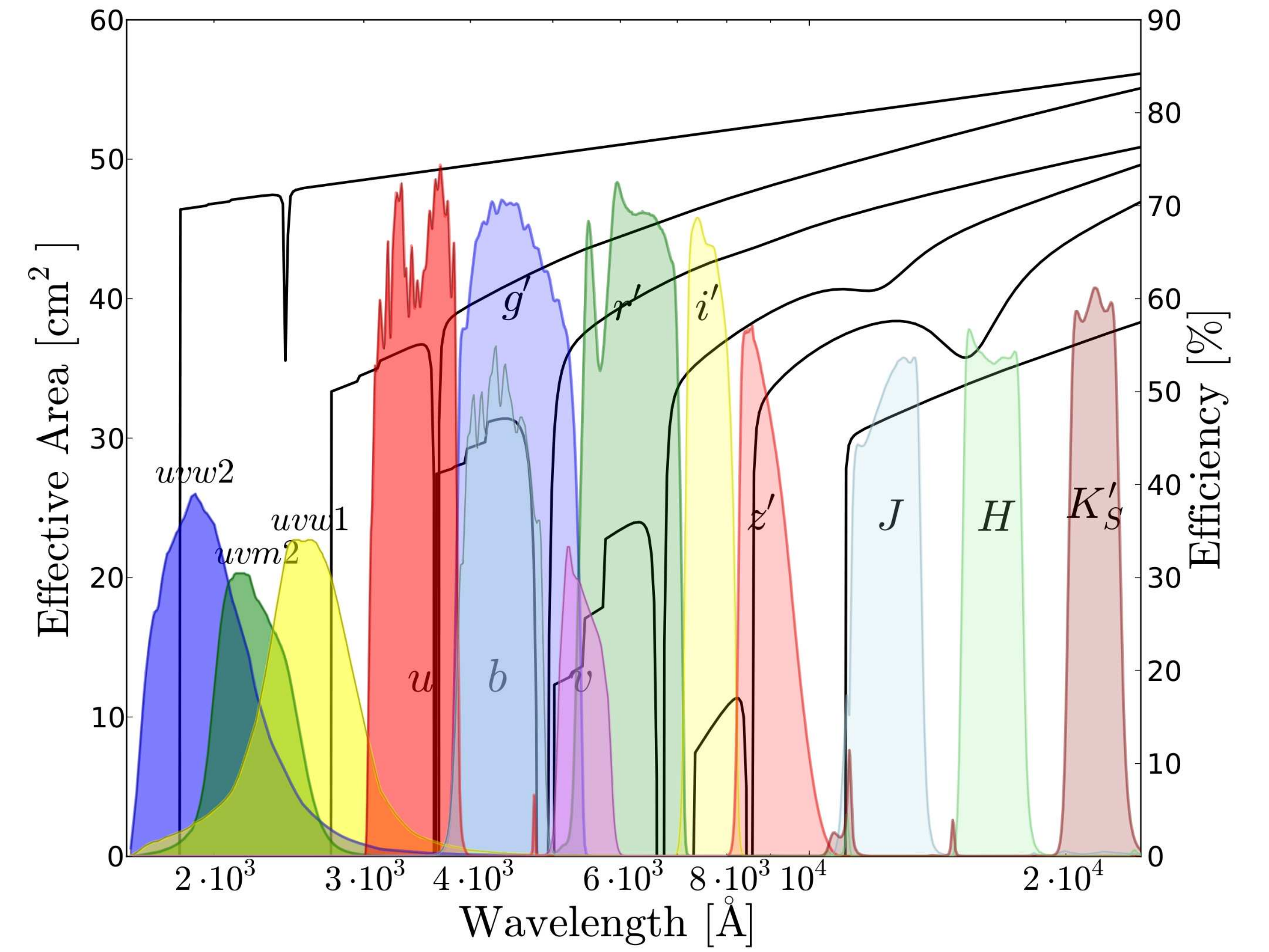}
\caption{Swift/UVOT ($uvw2$, $uvm2$, $uvw1$, $u$, $b$, $v$) effective areas 
(left y-axis) and GROND (\gp\rp\ip\zp$JHK_s$) filter transmission curves 
(right y-axis), respectively. The GROND filter curves include all 
optical components including the telescope. Shown with black solid lines 
are template afterglow spectra for redshifts z = 1, 2, 3, 4.5, 6 to 8 
(top left to bottom right). These spectra also differ in their spectral index 
and rest-frame extinction (both, amount and reddening law),
2175 $\mathring{A}$ dust feature, and damped Ly-$\alpha$ absorption. 
[From \cite{ksg11}].
}
\label{GROND_UVOT-filters}
\end{figure}

With the majority of GRBs occurring at redshift $<$2, but the lowest-redshift
GROND dropouts (\gp\ dropout) measurable only at around $z \sim 3$, 
GROND alerts to the community for high-redshift GRBs was not particularly large,
at the 20\% rate. However, the sensitivity can be extended to lower
redshifts by combining GROND observations with simultaneous
observations at wavelengths bluer than GROND-\gp. For instance, the 
combination with Swift/UVOT (Fig. \ref{GROND_UVOT-filters}) allows 
photometric redshifts as low as $z \sim 1.2$. In this particular case,
even non-simultaneous observations can be used, as the UVOT-$b$ filter
closely matches the GROND-\gp\ filter, which thus can be used for the
relative cross-calibration.

\begin{figure}[t]
\hspace{-0.5cm}
\includegraphics[width=9.cm]{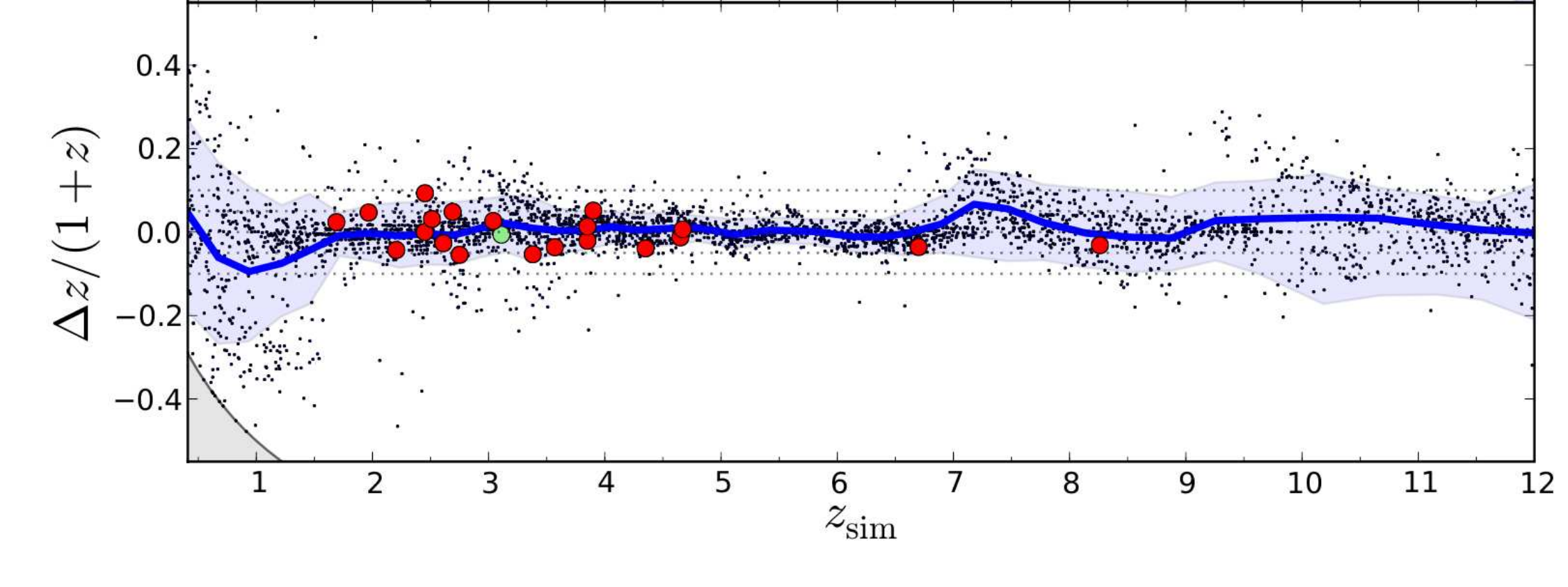}
\caption{Simulated photometric redshift accuracy (grey dots) vs. observed
GRBs with GROND+Swift/UVOT for which spectroscopic measurements are available 
(red dots).
The green dot shows the photo-z of the flat-spectrum radio quasar PKS 0537-286
derived in a similar manner (z = 3.10; \cite{Bottacini+2010}).
The thick blue line shows the average photometric redshift after distributing 
the 4000 mock afterglows into redshift bins of 100 afterglows each,
and the blue-shaded area shows the 1$\sigma$  statistical uncertainty.
[From \cite{ksg11}].
}
\label{z-accuracy}
\end{figure}

Examples for such application are GRB afterglows \citep{ksg11}
and BL Lac objects \citep{Rau+2012, Kaur+2017}.
In both cases, the intrinsic spectral energy distribution is a (sometimes 
broken) power law, and thus the Ly-$\alpha$ drop creates a clear signature,
leading to typical photometric redshifts errors of 
$\Delta z / (1+z) \sim 10\%$ for $z > 1.5$ 
(Fig. \ref{z-accuracy}).
The dust-redshift degeneracy is broken with increasing redshift, as the 
Ly-limit moves to redder wavelengths, producing a drop-out at blueer filters 
which is too sharp to be mistaken by dust absorption. 
The redshift accuracy remains essentially constant until 
$z \sim 6.5$, demonstrating that the total number of individual filters 
does not strongly affect the robustness of the photo-z measurement, as long as 
the intrinsic continuum is fairly well known (as in GRBs and BL Lacs). 

\section{High-redshift quasars}

\begin{figure}[t]
\includegraphics[width=8.5cm]{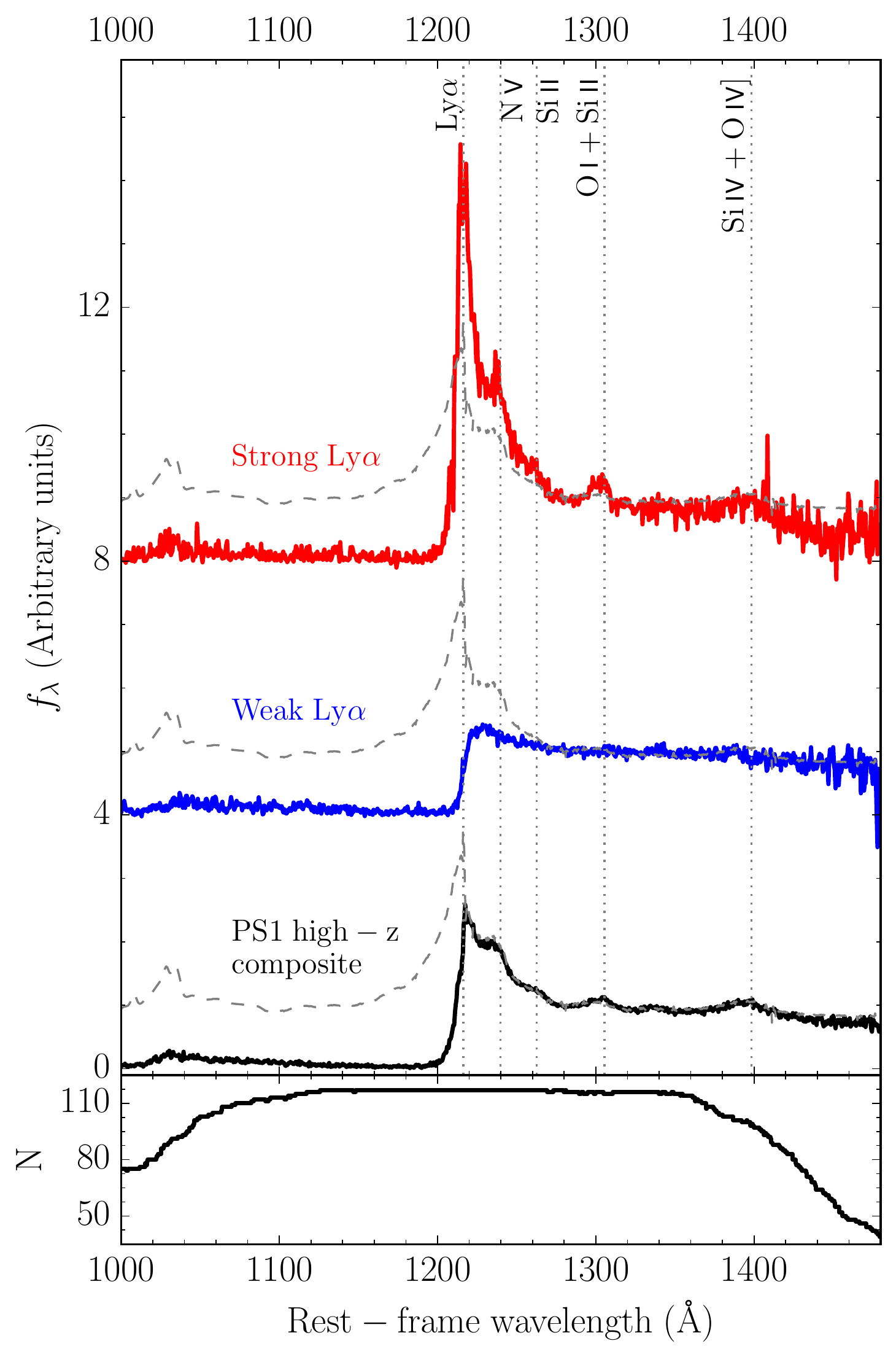}
\caption{Top: Diversity of high-z QSO emission spectra, showing those with
the 10\% strongest (red) and weakest (blue) Ly-$\alpha$ equivalent width,
compared to the low-redshift composite (gray) of \cite{Selsing+2016}
and the arithmetic median of all 117 QSOs (black) of \cite{Banados+2016}.
Bottom: number of QSOs per wavelength bin contributing to the median
of 117 QSO spectra.
[From \cite{Banados+2016}].
}
\label{Lyalpha}
\end{figure}

The search for high redshift ($z$ \gax\ 6) has been an area of intense work
over the last decade, given that early versions of their luminosity functions
indicated rather high surface density. The majority of searches were based
on the dropout technique, and the challenge for color selections is the
high incident of low-redshift contaminating sources (e.g. brown dwarfs,
redshift two galaxies). Down-selecting candidate lists in preparation of
follow-up spectroscopy then is the area where imaging with GROND is most 
efficient. Indeed, GROND has played a major role in this task
\citep{Banados+2014, Venemans+2015, Banados+2016, Mazzucchelli+2017}.

After achieving a sample size of a few dozen quasars at $z>6$, several 
physically interesting new aspects arose. One of those, with direct impact on
multi-colour imaging, is the unexpectedly large variance in the strength
of the Ly-$\alpha$ (+ \ion{N}{5}) emission line (see Fig. \ref{Lyalpha}):
for the 10\% of quasars with the smallest Ly-$\alpha$ + \ion{N}{5} equivalent 
width, the line is virtually absent, and thus they resemble weak-emission line 
objects \citep{Banados+2016}. This is substantially more than at lower
redshift. But whether this is an evolutionary effect or an 
observational bias (by the filter setting or the increase in  the neutral
hydrogen fraction, or both), remains to be investigated.

\section{Conclusion}

The design of the GROND instrument was originally developed for GRB afterglow 
observations. Nevertheless, many other science areas have greatly benefited 
from the simultaneous 7-channel imaging of GROND. GRB follow-up 
observations used only about 15\% of the 2.2m telescope time between 2008--2016,
yet provided noteworthy results for 112 refereed publications, and
supported 12 PhD theses. The biggest impact of GROND observations of GRBs
was undoubtly the initiation of systematic afterglow follow-up in the
near-infrared ($JHK_s$), enabling (i) the discovery of high-redshift GRBs
as well as (ii) studying the prevalence of dust along GRB sightlines, i.e. 
quantifying the incidence of 'dark' bursts.
Other scientific highlights include 
(iii) studies of Fermi-detected GRBs
(including the redshift estimate of GRB 080916C), 
(iv) measurements of the SEDs of non-canonical light curve variability such
  as flares or intensity jumps,
(v) the discovery and detailed study of a large fraction of all GRB-SNe,
  most prominently the SN 2011kl related to the ultra-long GRB 111209A,
(vi) tests of the simplest fireball scenario based on the evolution of
  afterglow SEDs,
(vii) the investigation of the jet structure and/or off-axis viewing geometry
  based on achromatic afterglow light curves,
(viii) and the characterization of about half of all optically-detected 
short GRB afterglows.
Beyond GRBs, the black widow binaries and photometric redshift estimates
for blazars are the most noteworthy topics.

Over the years, the versatility of GROND has made it the instrument with the
largest share of observing time among the three instruments at the 2.2m ESO/MPG 
telescope. This has been a rewarding experience for the team that
designed and built the instrument. It also demonstrates
that specialized instruments with unique capability at a small telescope can 
indeed make competitive contributions in the era of 8-10\,m telescopes,
and likely also in the upcoming era of even larger telescopes.

\bigskip

\acknowledgments
\noindent{\it I particularly acknowledge U. Laux for the mastery of the 
optical design of GROND, and S. Klose (both Th\"uringer Landessternwarte 
Tautenburg) for the long-standing fruitful
collaboration since the early time of the development of GROND,  
as well as T. Kr\"uhler (formerly MPE) for his
unprecedented breadth of technical and scientific insight, 
which substantially contributed to shape the success of GROND.
I'm grateful for the enthusiasm and help of all the GRB-GROND team members 
over the years: P.M.J. Afonso, J. Bolmer, C. Clemens, C. Delvaux, J. Elliott,
R. Filgas, J.F. Graham, D.A. Kann, F. Knust, A. K\"upc\"u Yolda\c{s},  
M. Nardini, A. Nicuesa Guelbenzu, F. Olivares E., N. Primak, A. Rossi,
P. Schady, S. Schmidl, T. Schweyer, G. Szokoly, I. Steiner, V. Sudilovsky, 
M. Tanga, C.C. Th\"one, K. Varela, P. Wiseman, and A. Yolda\c{s}.
I appreciate rewarding discussions with D.H. Hartmann, M. Ajello, B. Stecklum,
H. van Eerten, L. Mancini, B. Biller, 
and E. Banados, and the always instant support by the ESO La Silla crew
in all circumstances.
Paulo M.J. Afonso (now at the American River College, U.S.A.)
identified the very red object described in sect. 6.3, leading
to the VLT follow-up spectroscopy.
Part of the funding for GROND (both hardware as well as personnel)
was generously granted from the Leibniz-Prize (DFG grant HA 1850/28-1)
to Prof. G. Hasinger. Additional funding was provided by the 
Th\"uringer Landessternwarte Tautenburg.}

\bigskip

\facilities{Max Planck:2.2m, GROND instrument}.


\bigskip


\begin{thebibliography}{}

\bibitem[Abbott et al.(2017a)]{LIGO+2017} Abbott B.P., Abbott R., Abbott T.D.
et al. 2017a, PRL 119, 161101

\bibitem[Abbott et al.(2017b)]{LIGO_APJL_MM} Abbott B.P., Abbott R., Abbott T.D.
et al. 2017b, ApJ,L 848, L12

\bibitem[Abdollahi et al.(2017)]{Abdollahi+2017} Abdollahi S., Ackermann M.,
Ajello M. et al. 2017, ApJ, 846, 34

\bibitem[Akitaya et al.(2014)]{Akitaya2014} Akitaya H., Moritani Y.,
Ui T., et al. 2014, SPIE 9147, 91474

\bibitem[Alexander et al.(2017)]{Alexander+2017} Alexander K.D., Laskar T.,
Berger E., et al. 2017, ApJ, 848, 69

\bibitem[Allard et al.(2012)]{Allard+2012} Allard F., Homeier D., \& Freytag B.,
2012, RSPTA, 370, 2765

\bibitem[Aloy et al.(2005)]{Aloy+2005} Aloy M.A., Janka H.-T., \& M\"uller E.,
2005, A\&A 436, 273

\bibitem[Banados et al.(2014)]{Banados+2014} Banados E., Venemans B.P.,
Morganson E., et al. 2014, AJ, 148, 14

\bibitem[Banados et al.(2016)]{Banados+2016} Banados E., Venemans B.P.,
Decarli R., et al. 2016, ApJS 227, 11

\bibitem[Beletic et al.(1998)]{Beletic1998} Beletic J.W., Gerdes R., \&
Duvarney R.C., 1998, Proc. ESO CCD workshop,  Garching, 8-10 Oct. 1996,
 Eds. J.W. Beletic and P. Amico, Kluwer, ASSL 228, p. 103

\bibitem[Bernardes et al.(2018)]{Bernardes2018} Bernardes D.V., 
Martioli E., Rodrigues C.V., 2018, PASP 130, pp. 095002

\bibitem[Biller et al.(2013)]{Biller+2013} Biller B., Crossfield I.J.M., 
Mancini L. et al. 2013, ApJ, 778, L10

\bibitem[Bottacini et al.(2010)]{Bottacini+2010} Bottacini E., Ajello M., 
Greiner J. et al. 2010, A\&A, 509, A69

\bibitem[Burgasser et al.(2006)]{Burgasser+2006} Burgasser A.J.,
Geballe T.R., Leggett S.K. et al. 2006, ApJ, 637, 1067

\bibitem[Burgess et al.(2017)]{Burgess+2017} Burgess J.M., Greiner J.,
Begue D., et al., 2017, arXiv:1710.05823

\bibitem[Burningham et al.(2011)]{Burningham+2011} Burningham B., Leggett S.K.,
  Homeier D. et al. 2011, MN 414, 3590

\bibitem[Butler et al.(2012)]{Butler2012} Butler N., Klein C., Fox O.
et al. 2012, SPIE 8446, 844610

\bibitem[Cameron(2016)]{Cameron2016} Cameron A.C., 2016, in ``Methods of 
Detecting Exoplanets'', ASSL 428, Springer, 89

\bibitem[Cano et al.(2014)]{Cano+2014} Cano Z., de Ugarte Postigo A., 
Pozanenko A., et al. 2014, A\&A, 568, A19

\bibitem[Cano et al.(2017)]{Cano+2017} Cano Z., Wang S.-Q., Dai Z.-G., \&
Wu X.-F., 2017, Adv. Astron., ID 8929054 

\bibitem[Cappelluti et al.(2009)]{Cappelluti+2009} Cappelluti N., Ajello M.
Rebusco P., et al. 2009,  A\&A, 495, L9

\bibitem[Cenko et al.(2010)]{Cenko10} Cenko S.B., Frail D.A., Harrison F.A.,
  et al. 2010, ApJ, 711, 641

\bibitem[Cenko et al.(2011)]{Cenko11} Cenko S.B., Frail D.A., Harrison F.A.,
  et al. 2011, ApJ, 732, 29

\bibitem[Cenko et al.(2012)]{Cenko+2012} Cenko S.B., Krimm H.A., 
Horesh A., et al. 2012, ApJ, 753, 77

\bibitem[Chabrier et al.(2014)]{Chabrier+2014} Chabrier G., Johansen A., 
Janson M., \& Rafikov R., 2014, in ``Protostars and Planets VI'', 
eds. H. Beuther et al., Univ. Arizona Press, p. 619

\bibitem[Chandra et al.(2008)]{Chandra2008} Chandra, P., Cenko, S.B., Frail, 
D.A., et al.\ 2008, \apj, 683, 924

\bibitem[Chen et al.(2014)]{Chen+2014} Chen G., van Boekel R., Madhusudhan N.
et al. 2014, A\&A, 564, A6

\bibitem[Chrimes et al.(2018)]{Chrimes18} Chrimes A., Stanway E., Levan A.
et al. 2018, MN, 478, 2

\bibitem[Christille et al.(2016)]{Christille+2016} Christille J.M., 
Bonomo A.S., Borsa F. et al. 2016, SPIE 9908, id. 990857

\bibitem[Connelley et al.(2013)]{Connelley2013} Connelley M., Tokunaga A.,
Bus S., 2013, AAS DPS meeting \#45, id. 211.12

\bibitem[Costa et al.(1997)]{Costa1997} Costa E., Frontera F., Heise J., 
et al., 1997, Nat., 387, 783

\bibitem[Crossfield et al.(2014)]{Crossfield+2014} Crossfield I.J.M.,
Biller B., Schlieder J.E. et al. 2014, Nat., 505, 654

\bibitem[Cucchiara et al.(2011)]{Cucchiara+2011} Cucchiara A., Levan A.J., 
Fox D.B. et al. 2011, ApJ, 736, 7

\bibitem[Depoy(1998)]{dep98} Depoy D., 1998, URL
www.astronomy.ohio-state.edu/\~\,depoy/research/instrumentation/ andicam/andicam.html

\bibitem[Dhillon et al.(2007)]{dhi07} Dhillon V.S., Marsh T.R., Stevenson M.J.,
 et al., 2007, MN 378, 825

\bibitem[Djorgovski et al.(2001)]{dfk01} Djorgovski S.G., Frail D.A., 
  Kulkarni S.R., et al. 2001, ApJ, 562, 654

\bibitem[Dunham et al.(2004)]{dun04} Dunham E.W., Elliot J.L., Bida T.A., 
 et al. 2004, SPIE 5492, 592

\bibitem[Eliasd\'ottir et al.(2009)]{Eliasdottir+2009} Eliasd\'ottir \'A., 
Fynbo J.P.U., Hjorth J., et al.  2009, ApJ, 697, 1725

\bibitem[Elliott et a.(2012)]{Elliott+2012} Elliott J., Greiner J., 
Khochfar S., 
et al., 2012, A\&A, 539, A113

\bibitem[Ellis et al.(1993)]{edf92} Ellis T.A., Drake R., Fowler A.M.,
et al.  1993, in ``Cryogenic Optical Systems and Instruments V'',
 Proc. SPIE 1765, 94

\bibitem[Fan et al.(2013)]{Fan+2013} Fan Y.-Z., Yu Y.-W., Xu D., et al. 2013,
 ApJ, 779, L25 

\bibitem[Filgas et al.(2011a)]{Filgas+2011} Filgas R., Kr\"uhler T., Greiner J.
et al. 2011a, A\&A, 526, A113

\bibitem[Filgas et al.(2011b)]{fgs11} Filgas R., Greiner J., Schady P.
et al. 2011b, A\&A, 535, A57

\bibitem[Filgas et al.(2012)]{fgs12} Filgas R., Greiner J., Schady P.
et al. 2012, A\&A, 546, A101

\bibitem[Fox et al.(2005)]{Fox+2005} Fox D.B., Frail D.A., Price P.A. et al.
2005, Nat., 437, 845

\bibitem[Fruchter et al.(2006)]{Fruchter+2006} Fruchter A.S., Levan A.J.,
Strolger L. et al. 2006, Nat., 441, 463

\bibitem[Fukugita et al.(1996)]{fig96} Fukugita M., Ichikawa T., Gunn J.E.
 \etal\ 1996, AJ, 111, 1748

\bibitem[Fynbo et al.(2001)]{fyn01} Fynbo J.U., Jensen B.L., Gorosabel J. 
  et al. 2001, A\&A, 369, 373

\bibitem[Galama et al.(1998)]{Galama+1998} Galama T.J., Vreeswijk P.M., 
van Paradijs J. et al. 1998, Nat., 395, 670

\bibitem[Gehrels et al.(2004)]{Gehrels+2004} Gehrels N., Chincarini G., 
Giommi P.,  et al. 2004,  ApJ, 611, 1005

\bibitem[Gendre et al.(2013)]{gsa13} Gendre B., et al. 
ApJ, 766, A30

\bibitem[Goldman et al.(2010)]{Goldman+2010} Goldman B., Marsat S.,
Henning T., Clemens C., \& Greiner J., 2010, MN, 405, 1140

\bibitem[Golenetskii et al.(2011)]{Golen11} Golenetskii S., et al.
GCN Circ. 12663, http://gcn.gsfc.nasa.gov/gcn/gcn3/12663.gcn3

\bibitem[Gorosabel \& Ugarte Postigo(2010)]{Gorosabel2010} Gorosabel J., 
Ugarte Postigo A., 2010, in Proc. of 
``High Time Resolution Astrophysics IV - The era of extremely large 
telescopes'', Crete, May 2010, PoS 108, id. 36

\bibitem[Graham \& Fruchter(2017)]{Graham+Fruchter2017} Graham J.F., \&
Fruchter A.S., 2017, ApJ, 834, 170

\bibitem[Granot et al.(2002)]{Granot2002} Granot J., Panaitescu A,
Kumar P., \& Woosley S.E., 2002, ApJ, 570, L61

\bibitem[Granot \& Kumar(2003)]{GranotKumar03} Granot J., \& Kumar P., 2003, 
ApJ, 591, 1086 

\bibitem[Greiner et al.(2007)]{gck07a} Greiner J., Clemens C., Kr\"uhler T., 
\etal\ 2007, GCN Circ. 6449, http://gcn.gsfc.nasa.gov/gcn/gcn3/6449.gcn3

\bibitem[Greiner et al.(2008)]{Greiner+2008} Greiner J., Bornemann W., 
Clemens C., et al. 2008, PASP, 120, 405

\bibitem[Greiner et al.(2009a)]{gkf09} Greiner J., Kr\"uhler T., 
Fynbo J.P.U, et al. 2009a, ApJ, 693, 1610

\bibitem[Greiner et al.(2009b)]{gkm09} Greiner J., Kr\"uhler T., 
McBreen S., et al. 2009b, ApJ, 693, 1912

\bibitem[Greiner et al.(2011)]{gkk11}  Greiner J.,  Kr\"uhler T., Klose S. 
et al., 
2011, A\&A, 526, A30

\bibitem[Greiner et al.(2013)]{gkn13} Greiner J., Kr\"uhler T., Nardini M. 
et al. 2013, A\&A, 560, A70

\bibitem[Greiner et al.(2015)]{Greiner+2015} Greiner J., Mazzali P.A.,
Kann D.A. et al. 2015, Nat., 523, 189

\bibitem[Greiner et al.(2016)]{Greiner+2016} Greiner J., Delvaux C.,
Wyrzykowski L., et al. 2016, Astron. Tel. \#8579

\bibitem[Greiner et al.(2018)]{gbw18} Greiner J., Bolmer J., Wieringa M.
et al. 2018, A\&A, 614, A29

\bibitem[Groot et al.(1998)]{ggp98} Groot P.J., Galama T.J., van Paradijs J., 
  et al. 1998, ApJ, 493, L27

\bibitem[Gvaramadze et al.(2017)]{Gvaramadze+2017} Gvaramadze V.V., Langer N.,
Fossati L., et al. 2017, Nat. Astron., 1, 0116

\bibitem[Hjorth et al.(2003)]{Hjorth03} Hjorth J., Sollermann J., Moller P.
 et al. 2003, Nat., 423, 847

\bibitem[Inserra(2013)]{Inserra13} Inserra, C.,
2013, ApJ, 770, A128

\bibitem[Jewitt(2002)]{jew02} Jewitt D.C., 2002, AJ, 123, 1039

\bibitem[Jha et al.(2000)]{jha00} Jha S., Charbonneau D., Garnavich P.M. 
et al. 2000, ApJ, 540, L45

\bibitem[Kann et al.(2018)]{Kann+2016} Kann D.A., Schady P., Olivares E. F.,
et al. A\&A 617, A122

\bibitem[Kasen \& Bildsten(2010)]{KasBil10} Kasen, D., \& Bildsten, L.,
ApJ, 717, 245

\bibitem[Kasen et al.(2013)]{Kasen+2013} Kasen D., Badnell N. R., \&
Barnes J., 
2013, ApJ, 774, 25

\bibitem[Kaur et al.(2017)]{Kaur+2017} Kaur A., Rau A., Ajello M., et al. 2017,
ApJ, 834, 41

\bibitem[Katz(1994)]{Katz1994} Katz J.I., 1994, ApJ, 422, 248

\bibitem[Kennea(2015)]{Kennea2015} Kennea J.A., 2015, JHEAp, 7, 105

\bibitem[Kirkpatrick et al.(2011)]{Kirkpatrick+2011} Kirkpatrick J.D.,
Cushing M.C., Gelino C.R. et al. 2011, ApJS, 197, 19 

\bibitem[Kleiser et al.(2011)]{Kleiser11} Kleiser I.K.W., et al. 2011,
MN, 415, 372

\bibitem[Klose et al.(2000)]{ksm00} Klose S., Stecklum B., Masetti N.,
et al. 2000, ApJ, 545, 271

\bibitem[Klose et al.(2018)]{Klose+2018} Klose S., Schmidl S., Kann D.A.
et al. 2018, A\&A, in press [arXiv:1808.02710]

\bibitem[Knust et al.(2017)]{Knust+2017} Knust F., Greiner J., 
van Eerten H., et al. 2017, A\&A, 607, A84

\bibitem[Komossa et al.(2009)]{Komossa+2009} Komossa S., Zhou H., Rau A.,
et al. 2009, ApJ, 701, 105 

\bibitem[Komossa(2017)]{Komossa2017} Komossa S., 2017, AN 338, 256

\bibitem[Kotani et al.(2005)]{kky07} Kotani T., Kawai N., Yanagisawa K., 
et al. 2005, Il Nuovo Cim. 28 C, p. 755 (astro-ph/0702708)

\bibitem[Kr\"uhler et al.(2008)]{kkg08} Kr\"uhler T., K\"{u}pc\"{u} Yolda\c{s}
 A., Greiner J.,   et al. 2008, ApJ, 685, 376

\bibitem[Kr\"uhler et al.(2009a)]{kgm09} Kr\"uhler T., Greiner J., McBreen S., 
et al. 2009a, ApJ, 697, 758

\bibitem[Kr\"uhler et al.(2009b)]{kga09} Kr{\"u}hler T., Greiner J., 
Afonso P., et al.\ 2009b, A\&A, 508, 593

\bibitem[Kr\"uhler et al.(2011a)]{ksg11} Kr\"uhler T., Schady P., Greiner J.
et al. 2011a, A\&A, 526, A153

\bibitem[Kr\"uhler et al.(2011b)]{kgs11} Kr\"uhler T., Greiner J., Schady P.,
et al. 2011b, A\&A, 531, A108

\bibitem[Kumar \& Piran(2000)]{Kumar+Piran2000} Kumar P., \& Piran T., 2000, 
ApJ, 535, 152

\bibitem[Lamb \& Reichart(2000)]{lar00} Lamb D.Q., \& Reichart D.E., 2000, 
ApJ, 536, 1

\bibitem[Laskar et al.(2014)]{Laskar14} Laskar T., Berger E., Tanvir N.R.,
et al. 2014, ApJ, 781, 1

\bibitem[Lattimer \& Prakash(2011)]{Lattimer+2011} Lattimer J.M., \&
Prakash M., 2011, in ``From Nuclei to Stars'', ed. S. Lee, World Scientific, 
p. 275

\bibitem[Lazzati, Covino, Ghisellini(2002)]{lcg02} Lazzati D., Covino S., 
 \& Ghisellini G., 2002a, MN, 330, 583

\bibitem[Lazzati et al.(2002)]{Lazzati2002} Lazzati D., Rossi E., Covino S., 
  Ghisellini G., \& Malesani D., 2002, A\&A, 396, L5

\bibitem[Le \& Mehta(2017)]{Le+Mehta17} Le T., \& Mehta V., 2017, ApJ, 837, 17

\bibitem[Lendl et al.(2010)]{Lendl+2010} Lendl M., Afonso C., Koppenhoefer J.
et al. 2010, A\&A, 522, A29

\bibitem[Levan et al.(2014)]{lts14} Levan A., et al.
ApJ, 781, A13

\bibitem[Lipunov et al.(2001)]{Lipunov+2001} Lipunov V.M., Postnov K.A., \&
Prokhorov M.E., 2001, Astron. Rep. 45, 236

\bibitem[Lodato et al.(2015)]{Lodato+2015} Lodato G., Franchini A.,
Bennerot C., \& Rossi E.M., 2015, JHEAp 7, 158

\bibitem[Luhman(2013)]{Luhman2013} Luhman K.L., 2013, ApJ, 767, L1

\bibitem[Lunnan et al.(2013)]{Lunnan13} Lunnan, R., et al.
2013, ApJ, 771, A97

\bibitem[Mancini et al.(2013)]{Mancini+2013} Mancini L., Nikolov N.,
Southworth J., et al. 2013, MN, 430, 2932

\bibitem[Mancini et al.(2014)]{Mancini+2014} Mancini L., Southworth J., 
Ciceri S., et al. 2014, A\&A, 568, A127

\bibitem[Mazzucchelli et al.(2017)]{Mazzucchelli+2017} Mazzucchelli C.,
Banados E., Venemans B.P., et al. 2017, ApJ, 849, 91

\bibitem[McBreen et al.(2010)]{mkr10}  McBreen S., Kr\"uhler T., 
   Rau A. et al. 2010, A\&A, 516, A71

\bibitem[Mereghetti et al.(2010)]{Mereghetti+2010} Mereghetti S., Paizis A., 
G\"otz D., et al. 2010, GCN Circ. 10555, http://gcn.gsfc.nasa.gov/gcn/gcn3/10555.gcn3

\bibitem[Merloni et al.(2015)]{Merloni+2015} Merloni A., Dwelly T., Salvato M.,
et al. 2015, MN, 452, 69 

\bibitem[Meszaros \& Rees(1997)]{mer97} Meszaros P., \& Rees M.J., 
1997, ApJ, 476, 232

\bibitem[Metzger et al.(2010)]{Metzger+2010} Metzger B.D., Martinez-Pinedo G.,
 Darbha S., et al. 2010, MN, 406, 2650

\bibitem[Meyer et al.(1998)]{Meyer1998} Meyer M., Finger G., Mehrgan H., 
et al. 
1998, in Proc. ``Infrared Astronomical Instrumentation'', ed. A.M. Fowler,
 SPIE Vol. 3354, p. 134

\bibitem[Mohler-Fischer et al.(2013)]{Mohler-Fischer+2013} Mohler-Fischer M.,
Mancini L., Hartman J.D., et al. 2013, A\&A,  558, A55

\bibitem[Mookerjea et al.(2010)]{Mookerjea+2010} Mookerjea B., Parisi P.,
Bhattacharyya S., et al. 2010, MN, 409, L114

\bibitem[Nakauchi et al.(2013)]{nks13} Nakauchi D., Kashiyama K., Suwa Y., \&
Nakamura T.,
ApJ, 778, A67 

\bibitem[Nardini et al.(2011)]{Nardini+2011} Nardini M., Greiner J., 
Kr\"uhler T., et al. 2011, A\&A, 531, A39

\bibitem[Nardini et al.(2014)]{Nardini+2014} Nardini M., Elliott J., Filgas R.,
et al. 2014, A\&A, 562, A29

\bibitem[Negoro(2017)]{Negoro2017} Negoro H. and MAXI team, 2017, in
``7 years of MAXI'', Proc. conf. Dec. 2016, p. 15 
(online at http://maxi.riken.jp/conf/sevenyears/pdf/O\_03.pdf)

\bibitem[Nicuesa Guelbenzu et al.(2011)]{NicuesaGuelbenzu+2011} Nicuesa 
Guelbenzu A., Klose S., Rossi A., et al. 2011, A\&A, 531, L6

\bibitem[Nicuesa Guelbenzu et al.(2012)]{NicuesaGuelbenzu+2012} Nicuesa 
Guelbenzu A., Klose S., Greiner J., et al. 2012, A\&A, 548, A101

\bibitem[Olivares et al.(2012)]{Olivares+2012} Olivares Estay F., Greiner J., 
Schady P. et al. 2012, A\&A, 539, A76

\bibitem[Olivares et al.(2015)]{Olivares+2015} Olivares Estay F., Greiner J., 
Schady P. et al. 2015, A\&A, 577, A44 

\bibitem[Orosz \& Bailyn(1997)]{orosz} Orosz J.A., \& Bailyn C.D., 1997, 
  ApJ, 477, 876

\bibitem[Paczynski(1986)]{pac86} Paczynski B., 1986, ApJ, 304,1

\bibitem[Paczynski \& Rhoads(1993)]{PaczynskiRhoads1993} Paczynski B., 
\& Rhoads J., 1993, ApJ, 418, L5

\bibitem[Panaitescu et al.(1998)]{Panaitescu1998} Panaitescu A.,
M\'esz\'aros P., \& Rees M., 1998, ApJ, 503, 314

\bibitem[Panaitescu \& Kumar(2002)]{PanaiKumar02} Panaitescu A., \& Kumar P., 
2002, ApJ, 571, 779

\bibitem[Pian et al.(2006)]{Pian+06} Pian, E. et al.
2006, Nat., 442, 1011

\bibitem[Pierini et al.(2012)]{Pierini+2012} Pierini D., Suhada R.,
Fassbender R., et al. 2012, A\&A, 540, A45

\bibitem[P2PP Manual(2007)]{p2pp} P2PP (Phase2 Proposal Preparation) Manual, 2007,
v. 2.13, Issue 9, Doc. No. VLT-MAN-ESO-19200-1644

\bibitem[Rau et al.(2012)]{Rau+2012} Rau A., Schady P., Greiner J. et al. 2012,
 A\&A, 538, A26

\bibitem[Rau(2012)]{Rau2012} Rau A., extracted from 
http://www.mpe.mpg.de/$\sim$arau/GRBs\_bonn.pdf

\bibitem[Rees \& Meszaros(1998)]{Rees+Meszaros1998} Rees M.J., \& Meszaros P.,
 1998, ApJ, 496, L1

\bibitem[Reif et al.(1999)]{rei99} Reif K., Bagschik K., de Boer K.S.,
 et al., 1999, SPIE 3649, 109

\bibitem[Rhoads(1999)]{Rhoads99} Rhoads J.E., 1999, ApJ, 525, 737

\bibitem[Rodrigues et al.(2012)]{Rodrigues+2012} Rodrigues C.V., Taylor K.,
Jablonski F. et al. 2012, SPIE 8446, 844626

\bibitem[Romani et al.(2012)]{Romani+2012} Romani R.W., Filippenko A.V.,
Silverman J.M. et al. 2012, ApJ, 760, L36

\bibitem[Romani et al.(2015)]{Romani+2015} Romani R.W., Filippenko A.V.,
\& Cenko S.B., 2015, ApJ, 804, 115

\bibitem[Roming et al.(2006)]{Roming+2006} Roming P.W.A., Vanden Berk D., 
Pal'shin V., et al. 2006, ApJ, 651, 985 

\bibitem[Roming et al.(2018)]{Roming2018} Roming P.W.A., van der Horst A.,
OCTOCAM Team, 2018, AAS Meeting \#231, id. 314.06

\bibitem[Rossi et al.(2012)]{Rossi+2012} Rossi A., Klose S., Ferrero P.,
et al. 2012, A\&A, 545, A77

\bibitem[Salvetti et al.(2015)]{Salvetti+2015} Salvetti D., Mignani R.P.,
De Luca A. et al. 2015, ApJ, 814, 88

\bibitem[Salvetti et al.(2017)]{Salvetti+2017} Salvetti D., Mignani R.P.,
De Luca A. et al. 2017, MN, 470, 466

\bibitem[Sari \& Piran(1997)]{SariPiran97} Sari R., \& Piran T., 1997, 
ApJ, 485, 270

\bibitem[Sauer et al.(2006)]{Sauer06} Sauer, D.N. et al.
2006, MN, 369, 1939

\bibitem[Selsing et al.(2016)]{Selsing+2016} Selsing J., Fynbo J.P.U.,
Christensen L., \& Krogager J.-K., 2016, A\&A, 585, A87

\bibitem[Smartt et al.(2017)]{Smartt+2017} Smartt S.J., Chen T.-W., 
Jerkstrand A., et al. 2017, Nat., 551, 75

\bibitem[Snellen et al.(2009)]{Snellen+2009} Snellen I.A.G., 
Koppenh\"ofer J., van der Burg R.F.J., et al. 2009, A\&A, 497, 545

\bibitem[Spano et al.(2010)]{Spano2010} Spano P., De Caprio V., La Foresta M.
et al. 2010, Proc. SPIE 7735, id. 773565

\bibitem[Stamatellos(2017)]{Stamatellos2017} Stamatellos D., 2017, 
in ``Star formation from cores to clusters'', Proc. conf. Santiago,
March 2017, id. 56
(online at http://www.eso.org/sci/meetings/2017/star-formation2017.html)

\bibitem[Stanek et al.(2003)]{Stanek03} Stanek, K.Z., Matheson T., 
Garnavich P.M. et al. 2003, ApJ, 591, L17

\bibitem[Stone et al.(2018)]{Stone+2018} Stone N.C., Kesden M., Cheng R.M., \&
van Velzen S., 2018, Gen. Relativ. and Grav., submitted, [arXiv:1801.10180] 

\bibitem[Stratta et al.(2007)]{Stratta+2007} Stratta G., D'Avanzo P., 
Piranomonte S., et al. 2007, A\&A, 474, 827

\bibitem[Tanvir et al.(2009)]{Tanvir2009} Tanvir N.R., Fox D.B., Levan A.J. 
et al. 2009, Nat., 461, 1254

\bibitem[Taubenberger et al.(2006)]{tpm06} Taubenberger S., Pastorello A., 
Mazzali P.A., et al. 2006,
MN, 371, 1459

\bibitem[Teplitz et al.(2000)]{Teplitz+2000} Teplitz H.I., Malumuth E., 
Woodgate B.E. et al. 2000, PASP, 112, 1188

\bibitem[Updike et al.(2010)]{Updike+2010} Updike A., Rau A., Olivares E.F.,
\& Greiner J., 2010, GCN Circ. 10558, http://gcn.gsfc.nasa.gov/gcn/gcn3/10558.gcn3

\bibitem[van Paradijs et al.(1997)]{Paradijs1997} van Paradijs J., Groot P.J.,
 Galama T., et al., 1997, Nat., 386, 686

\bibitem[Varela et al.(2016)]{Varela+2016} Varela K., van Eerten H., 
Greiner J. et al., 2016, A\&A, 589, A37

\bibitem[Varela(2017)]{Varela2017} Varela K., 2017, PhD thesis,
TU Munich

\bibitem[Venemans et al.(2015)]{Venemans+2015} Venemans B.P., Banados E., 
Decarli R., et al. 2015, ApJ, 801, L1

\bibitem[Vlasis et al.(2011)]{vvm11} Vlasis A., van Eerten H.J., Meliani Z., 
\&  Keppens R., 2011, MN, 415, 279

\bibitem[Wang \& Loeb(2000)]{Wang+Loeb2000} Wang X., \& Loeb A., 2000, ApJ, 535, 788

\bibitem[Watanabe et al.(2005)]{wny05} Watanabe M., Nakaya H., Yamamuro T.,
\etal, 2005, PASP, 117, 870

\bibitem[Wijers et al.(1997)]{wij97} Wijers R.A.M.J., et al. 1997, MN, 288, L51

\bibitem[Wijers \& Galama(1999)]{wig99} Wijers R.A.M.J., \& Galama T., 
1999, ApJ, 523, 177

\bibitem[Woosley et al.(1999)]{Woosley+1999} Woosley S.E., Eastman R.G., \&
Schmidt B.P., 1999, ApJ, 516, 788

\bibitem[Wyrzykowski et al.(2016)]{Wyrzykowski+2016} Wyrzykowski L.,
Kostrzewa-Rutkowska Z., Udalski A., et al. 2016, Astron. Tel. \#8577 

\bibitem[Wyrzykowski et al.(2017)]{Wyrzykowski+2017} Wyrzykowski L.,
Zieli\'nski M., Kostrzewa-Rutkowska Z., et al. 2017, MN, 465, L114

\bibitem[Yates et al.(2015)]{Yates2015} Yates R., Kann D.A., Delvaux C., 
\& Greiner J., 2015, GCN Circ. 18674, 
http://gcn.gsfc.nasa.gov/gcn/gcn3/18674.gcn3

\bibitem[Yost et al.(2003)]{Yost03} Yost S.A., Harrison F.A., Sari R., \&
  Frail D.A., 2003, ApJ, 597, 459

\end{thebibliography}
\end{document}